\documentclass[12pt]{article}
\usepackage{amsfonts}
\usepackage{amssymb}
\usepackage{amsmath}
\usepackage{geometry}
\usepackage{color}
\usepackage[colorlinks=true,linkcolor=linkblue,citecolor=linkblue,urlcolor=linkblue]{hyperref}

\definecolor{linkblue}{rgb}{0,0,0.55}
\hypersetup{pdfborder={0 0 0}}
\newtheorem{theorem}{Theorem}

\newtheorem{corollary}{Corollary}

\newtheorem{definition}[theorem]{Definition}

\newtheorem{lemma}{Lemma}

\newtheorem{proposition}{Proposition}
\newtheorem{remark}{Remark}

\newenvironment{proof}[1][Proof]{\textbf{#1.} }{\ \rule{0.5em}{0.5em}}
\begin{document}

\title{Rating Multiple Agents with Career Concerns\thanks{%
This paper supersedes an earlier version titled `` Collective
vs. Relative Performance Evaluations with Career
Concerns.''\ I thank participants at the 2024 Bosphorus
Workshop on Economic Design and the Duke and UNC Theory Lunches for
comments, the Toulouse School of Economics and the University of St.~Gallen
for their hospitality, and Zhiyuan Jia, Xuerong Wang, and Runzhe (Bill) Zhao
for research assistance. All errors are mine.}}
\author{Huseyin Yildirim\medskip \\
Department of Economics\\
Duke University\\
Box 90097\\
Durham, NC 27708\medskip \\
E-mail: hy12@duke.edu}
\maketitle

\begin{abstract}
A principal designs ratings for multiple career-concerned agents with
correlated talents, mapping outputs into arbitrary finite-variance
statistics. Peer output both benchmarks effort and signals talent, making
full disclosure suboptimal under confidential ratings. Relaxing Bayes
plausibility to linear orthogonality yields a closed-form rating and
computable effort bounds for general priors; under normality, the relaxation
is exact. Optimal confidential ratings combine the full-disclosure wage with
the Fisher score for effort: they are relative-performance when individual
output is informative about talent and team-based when it is noisy. Public
ratings must satisfy joint Bayes plausibility, weakening incentives; hence,
the principal prefers confidential ratings, whereas agents prefer public
ratings. A spectral decomposition characterizes public ratings under
normality: small cost heterogeneity implies full disclosure regardless of
the noise structure, whereas large cost heterogeneity or strong effort
spillovers can induce partial disclosure to reallocate incentives. The
results bear on performance ratings and pay transparency.

\medskip


\smallskip


\newpage
\end{abstract}

\section{Introduction}

\label{sec:intro}

\setlength{\baselineskip}{18pt}Performance ratings shape incentives and careers by
affecting pay, promotion, and job opportunities. Yet their design remains a
challenge, as policy shifts across sectors illustrate. In the private
sector, Adobe, Microsoft, and General Electric moved away from relative
rankings by the mid-2010s to avoid penalizing top talent in elite cohorts
and encourage collaboration, whereas Google, Meta, and Amazon have since
2022 strengthened such practices to raise performance standards. In the
public sector, U.S. federal policy has traditionally prohibited forced
distributions, which cap how many employees may receive each rating, but
recent reforms have similarly sought greater differentiation, permitting
them for the broader civil service in 2026.\footnote{%
See Section~\ref{sec:discussion} for details and further discussion.} These
policy changes reflect a basic tension in how ratings should use peer
performance. This paper studies that tension when career
concerns---employment outcomes based on perceived talent---are the primary
source of incentives.

Career concerns are especially salient in government agencies, where formal
incentive pay is limited, but remain important across organizations.%
\footnote{%
In a recent Pew survey, 65 percent of employed U.S. adults described their
current job as either a career (50 percent) or a stepping stone to one (15
percent) (Lin et al. \hyperlink{cite.LinEtAl2024}{\textcolor{linkblue}{2024}}%
).} Because employees share hiring pools, training, and work conditions,
peer performance is informative about individual talent. This raises a key
design question: when peers excel, should the rating system reward an
employee because strong peers signal a talented cohort, or punish the
employee because his own performance is more attributable to talent than
effort? I study how a principal should optimally map performance data into
individual ratings to manage these reputational spillovers and motivate
agents.

The model is a familiar two-period version of Holmstr\"{o}m's (\hyperlink{cite.Holmstrom1999}%
{\textcolor{linkblue}{1999}}) career-concerns framework. A principal hires
multiple agents to perform separate tasks. Performance additively combines
unknown talent, unobservable effort, and noise. To isolate reputational
spillovers, I assume that peer output is `` good
news''\ about an agent's talent, with no effort spillovers
initially. Before agents simultaneously choose effort, the principal commits
to a rating system that maps the output vector into a finite-variance
statistic, including rank orders, total output, and multidimensional metrics.%
\footnote{%
Commitment can be achieved via institutional guidelines (e.g., HR policies)
for performance ratings.} The analysis allows general distributions before
specializing to a Gaussian environment for the full characterization. I
solve for the optimal rating system under two labor-market structures: 
\textbf{confidential ratings}, where each market observes only its own
agent's rating, as may occur across separate divisions, geographies, or
schools, and \textbf{public ratings}, where all markets observe the full
rating profile. Competitive markets (internal or external to the
organization) set wages equal to expected talent conditional on the
available rating information and perceived effort.

Incentives depend on the covariance between the agent's posterior wage given
the rating information and the Fisher score---the local likelihood
ratio---for effort. Strong peer performance thus plays a dual role: it
raises the market's assessment of an agent's talent but makes the agent's
own output less indicative of effort. Under confidential ratings, this
negative peer covariance makes full output disclosure suboptimal, since an
own-output rating would eliminate this effect. Talent similarity therefore
creates reputational free-riding and weakens incentives.

For general distributions, I relax Bayes plausibility to its linear
orthogonality requirement. This yields computable upper and lower bounds on
equilibrium effort, with the lower bound already improving on full
disclosure. Under Gaussianity, the bounds coincide, yielding a complete
characterization of the optimal rating and sharp comparative statics.

I call an agent's rating \textbf{team-based} if it rewards peer performance
on net, and \textbf{relative-performance} if it penalizes it. Under
independent noise, the optimal rating is team-based when the agent's own
output is sufficiently noisy, leveraging collective reputation concerns to
restore incentives, and relative-performance when own output is sufficiently
precise. Thus, heterogeneous signal-to-noise ratios can generate asymmetric
rating structures within the same organization. Correlated noise shifts
ratings toward relative performance, since it both weakens peer output as a
signal of talent and sharpens it as a benchmark for effort. Importantly, the
characterization reverses the implied firm policy: under full disclosure,
often assumed in the literature, a firm would favor hiring and training
practices that diversify coworkers' skills, whereas under optimal design it
would favor practices that homogenize them. Full disclosure overuses peer
output for talent inference, thereby exacerbating reputational free-riding.

Under public ratings, posterior beliefs must satisfy joint Bayes
plausibility: after observing the entire rating profile, no market should
revise its wage offer. Since every publicly feasible posterior is
confidentially feasible but not vice versa, publicity weakly reduces each
agent's equilibrium effort. The principal therefore prefers confidential
ratings, whereas agents prefer public ratings.\footnote{%
By iterated expectations, an agent's ex ante expected wages are equal across
all rating systems.} In the Gaussian environment, a spectral decomposition
fully characterizes the optimal public rating: it reveals exactly the
positive-eigenvalue components of the decorrelated full-disclosure
posterior. The number of components disclosed depends only on cost
parameters and the talent covariance matrix, not on the noise structure.%
\footnote{%
The approach resembles principal component analysis, but components are
selected by incentive value rather than explained variance.}

When cost heterogeneity is small, all components have positive incentive
value (i.e., positive associated eigenvalues), so full disclosure is
strictly optimal regardless of the noise structure, even when noise is
non-Gaussian. Public ratings are then team-based for every agent, whereas
confidential ratings may be relative.\footnote{%
Confidential ratings are independent of cost heterogeneity because they are
tailored to each agent without informational leakage.} When cost
heterogeneity is sufficiently large, partial disclosure becomes optimal: the
principal suppresses components with negative incentive value, limiting
reputational spillovers and redirecting incentives toward lower-cost agents.
Public ratings may then be less informative than confidential ratings about
some or all agents. Effort spillovers, positive or negative, play the same
role as cost heterogeneity: when agents' aggregate-talent exposures differ,
sufficiently strong spillovers can also make partial disclosure optimal.

Taken together, the results show that optimal ratings balance two uses of
peer performance: inferring talent and identifying individual effort.
Whether peer performance should be rewarded, penalized, or concealed depends
on both the information structure and the principal's incentive priorities.
As Section~\ref{sec:discussion} discusses in more detail, this perspective
informs ongoing debates over performance ratings. Forced distributions and
stack ranking may differentiate employees and intensify competition, but
peer comparisons can also suppress talent information that sustains career
incentives.\footnote{%
Stack ranking---popularized by GE's Jack Welch in the 1980s---forces
employees into fixed percentage tiers, traditionally a 20--70--10 split.}
Employer-enforced pay secrecy, and the legislative push against it, likewise
reflect the prediction that principals prefer confidential ratings while
agents prefer public ones.

\textbf{Related literature.} This paper builds on the canonical
career-concerns model of Holmstr\"{o}m (\hyperlink{cite.Holmstrom1999}{%
\textcolor{linkblue}{1999}}); see also Dewatripont et al. (\hyperlink{cite.DewatripontEtAl1999a}%
{\textcolor{linkblue}{1999a}},\hyperlink{cite.DewatripontEtAl1999b}{%
\textcolor{linkblue}{b}}). In these single-agent settings, the market
observes an exogenous performance signal, typically output. Recent work by
Rodina (\hyperlink{cite.Rodina2017}{\textcolor{linkblue}{2017}}) and H\"{o}%
rner and Lambert (\hyperlink{cite.HornerLambert2021}{%
\textcolor{linkblue}{2021}}) designs the disclosed signal to maximize effort.%
\footnote{%
Other single-agent models study rating design under dynamic reputations
(Ekmek\c{c}i \hyperlink{cite.Ekmekci2011}{\textcolor{linkblue}{2011}}),
coarse performance feedback (Hansen \hyperlink{cite.Hansen2013}{%
\textcolor{linkblue}{2013}}), intermittent monitoring of effort (Bar-Isaac
and Deb \hyperlink{cite.BarIsaacDeb2021}{\textcolor{linkblue}{2021}}),
dynamic evaluation and retention (Smolin \hyperlink{cite.Smolin2021}{%
\textcolor{linkblue}{2021}}), and strategic disclosure without commitment
(Pei \hyperlink{cite.Pei2016}{\textcolor{linkblue}{2016}}).} H\"{o}rner and
Lambert characterize optimal ratings in a fully dynamic Gaussian setting,
where the rating aggregates performance data of different vintages. I
consider a two-period setting with multiple agents, where the rating
aggregates the performance of peers, and allow general distributions through
a relaxation that is exact under Gaussianity.\footnote{%
Two periods is the minimal structure that generates reputational incentives
and is the workhorse framework in this literature, especially with multiple
agents. Section~\ref{sec:conclusion} discusses dynamics.} Full disclosure
remains optimal with a single agent, as in Rodina (\hyperlink{cite.Rodina2017}%
{\textcolor{linkblue}{2017}}), but is generally suboptimal with several,
because of reputational spillovers.

The paper also contributes to the multi-agent career-concerns literature.
Meyer and Vickers (\hyperlink{cite.MeyerVickers1997}{%
\textcolor{linkblue}{1997}}) show that peer comparisons strengthen or weaken
incentives depending on whether peer output mainly reflects common talent or
common noise. Their setting, with all outputs observed, corresponds to full
disclosure here; Sections~\ref{sec:char} and \ref{sec:public} further
develop this connection. Arya and Mittendorf (\hyperlink{cite.AryaMittendorf2011}%
{\textcolor{linkblue}{2011}}) compare aggregate and disaggregate performance
measures and find that aggregation can improve incentives. Meyer (\hyperlink{cite.Meyer1994}%
{\textcolor{linkblue}{1994}}), Auriol et al. (\hyperlink{cite.AuriolEtAl2002}%
{\textcolor{linkblue}{2002}}), Ortega (\hyperlink{cite.Ortega2003}{%
\textcolor{linkblue}{2003}}), Bar-Isaac (\hyperlink{cite.BarIsaac2007}{%
\textcolor{linkblue}{2007}}), Chalioti (\hyperlink{cite.Chalioti2016}{%
\textcolor{linkblue}{2016}}), Meagher and Prasad (\hyperlink{cite.MeagherPrasad2016}%
{\textcolor{linkblue}{2016}}), Neeman et al. (\hyperlink{cite.NeemanEtAl2019}%
{\textcolor{linkblue}{2019}}), and Yildirim (\hyperlink{cite.Yildirim2025}{%
\textcolor{linkblue}{2025}}) likewise study the incentive effects of career
concerns in teams. Unlike these papers, I optimize each agent's rating over
all finite-variance functions of the output vector and distinguish
confidential from public ratings. Neither full disclosure nor team output is
generally optimal.

More broadly, this paper relates to Bayesian persuasion (Kamenica and
Gentzkow \hyperlink{cite.KamenicaGentzkow2011}{\textcolor{linkblue}{2011}};
see Kamenica (\hyperlink{cite.Kamenica2019}{\textcolor{linkblue}{2019}}) for
a review). Under confidential ratings, each agent's design problem has the
product-of-posterior-means form of Rayo and Segal (\hyperlink{cite.RayoSegal2010}%
{\textcolor{linkblue}{2010}}), where the product arises from expected profit
times acceptance probability. Here it arises from effort incentives: the
posterior wage is paired with the Fisher score for effort, and the
multi-agent environment determines both through peer performance. Dworczak
and Kolotilin (\hyperlink{cite.DworczakKolotilin2024}{%
\textcolor{linkblue}{2024}}) develop a duality approach to this class of
multidimensional moment-persuasion problems. I instead relax Bayes
plausibility to linear orthogonality, yielding closed-form effort bounds for
general distributions and the exact optimum under Gaussianity. In
multi-receiver settings, Bardhi and Guo (\hyperlink{cite.BardhiGuo2018}{%
\textcolor{linkblue}{2018}}) compare signals based on all receivers' states
with signals based on a receiver's own state, and Arieli and Babichenko (
\hyperlink{cite.ArieliBabichenko2019}{\textcolor{linkblue}{2019}}) and Chan
et al. (\hyperlink{cite.ChanEtAl2019}{\textcolor{linkblue}{2019}}) compare
public with private disclosure, paralleling, respectively, the use of peer
and own performance and the two rating regimes here. Unlike these models,
disclosure in my setting motivates hidden effort. Aoyagi and Menuet (
\hyperlink{cite.AoyagiMenuet2026}{\textcolor{linkblue}{2026}}) similarly
design disclosure to a market that pays expected quality, but to elicit
private information from a single agent.

Finally, the paper relates to multi-agent moral hazard and
relative-performance evaluation. Under the informativeness principle, Holmstr%
\"{o}m (\hyperlink{cite.Holmstrom1982}{\textcolor{linkblue}{1982}})
establishes that peer output is valuable insofar as it filters common
shocks. This logic underlies classical models of relative-performance
evaluation and tournaments (Lazear and Rosen \hyperlink{cite.LazearRosen1981}%
{\textcolor{linkblue}{1981}}; Green and Stokey \hyperlink{cite.GreenStokey1983}%
{\textcolor{linkblue}{1983}}; Nalebuff and Stiglitz \hyperlink{cite.NalebuffStiglitz1983}%
{\textcolor{linkblue}{1983}}); see Fleckinger et al. (\hyperlink{cite.FleckingerEtAl2024}%
{\textcolor{linkblue}{2024}}) for a review. I abstract from explicit
contracts and instead examine career incentives generated by labor-market
learning. Absent career concerns, though, unknown talent would simply be
another production shock, again favoring relative-performance evaluation;
see, e.g., Holmstr\"{o}m (\hyperlink{cite.Holmstrom1982}{%
\textcolor{linkblue}{1982}}, Section 3). This literature typically assumes
public contracts. Halac et al. (\hyperlink{cite.HalacEtAl2021}{%
\textcolor{linkblue}{2021}}) show that confidential contracting improves
incentives through coordination rather than career concerns---a comparison I
return to in Section~\ref{sec:discussion}. Goldmanis and Ray (\hyperlink{cite.GoldmanisRay2021}%
{\textcolor{linkblue}{2021}}) and Buffa et al. (\hyperlink{cite.BuffaEtAl2025}%
{\textcolor{linkblue}{2025}}) instead highlight commitment problems created
by private contracting, which do not arise in my setting because wages are
competitively determined by the market rather than chosen by the principal
ex post.

The remainder of the paper proceeds as follows. Section~\ref{sec:model}
introduces the model. Section~\ref{sec:char} characterizes optimal
confidential ratings, and Section~\ref{sec:public} characterizes optimal
public ratings. Section~\ref{sec:spillovers} extends the model to effort
spillovers, Section~\ref{sec:discussion} discusses anecdotal evidence, and
Section~\ref{sec:conclusion} concludes.

\section{Baseline model: confidential ratings}

\label{sec:model}

I introduce performance ratings to a two-period\ version of Holmstr\"{o}m's (
\hyperlink{cite.Holmstrom1999}{\textcolor{linkblue}{1999}}) career-concerns
model. A principal\ (or organization)\ hires $n\geq 2$ agents on fixed
wages, normalized to zero. Agent $i$ exerts unobservable effort $a_{i}\in 
\mathbb{R}
_{+}$ and incurs a quadratic cost:\footnote{%
The quadratic cost helps obtain closed-form solutions but is not essential
for the qualitative results.}%
\begin{equation*}
c_{i}(a_{i})=\frac{a_{i}^{2}}{2\gamma _{i}}\text{.}
\end{equation*}%
The cost parameter $\gamma _{i}>0$ is publicly known, reflecting observable
features of agent $i$'s task or position, such as its scale or complexity.
Equivalently, agents may share the same cost function while the principal
weights agent $i$'s output by $\gamma _{i}$.

Agent $i$'s output is given by 
\begin{equation}
Y_{i}=\eta _{i}+a_{i}+\varepsilon _{i}\text{,}  \label{Output_i}
\end{equation}%
where $\eta _{i}\in 
\mathbb{R}
$ and $\varepsilon _{i}\in 
\mathbb{R}
$ denote his talent and exogenous noise, respectively. Writing vectors
without subscripts, let $\eta $ and $\varepsilon $ have means $\mu $ and $0$%
, and positive definite covariance matrices $\Sigma _{\eta }\succ 0$ and $%
\Sigma _{\varepsilon }\succ 0$, respectively, where $A\succ B$ signifies
that $A-B$ is positive definite (the L\"{o}wner order). Define base output $%
X\equiv \eta +\varepsilon $,\ so $Y=X+a$. Since incentives and Bayesian
inference will depend only on mean deviations, I henceforth center all
random variables and, without loss of generality, set $\mathbb{E}[\eta ]=%
\mathbb{E}[X]=0$.\footnote{%
Uncentered results are recovered by adding back the means; for example,
replace the centered $\eta $ by $\eta +\mu $.} To simplify notation, I
continue to use $\eta $ and $X$ for these centered variables.

Following the career-concerns literature, agents take actions uncertain of
their talents and noise terms, ruling out signaling motives. All players
share common priors over their distributions.

\noindent \textbf{\hypertarget{assm:1}{Assumption 1.}} \textup{%
(Informational environment)}

\begin{enumerate}
\item[(i)] $\eta $\textit{\ and }$\varepsilon $\textit{\ are independent
with strictly positive\ densities on }$\mathbb{R}^{n}$\textit{.}

\item[(ii)] \textit{The density }$f_{X}$\textit{\ is }$C^{2}$, \textit{%
log-concave, and log-supermodular.}

\item[(iii)] \textit{For every }$i$\textit{, }$\mathbb{E}[\eta _{i}$\textit{%
\ }$|$\textit{\ }$X]$\textit{\ is increasing in each component of }$X$%
\textit{.}
\end{enumerate}

Part (i) ensures regularity. Parts (ii) and (iii) drive the peer tension
studied below: higher peer output lowers the likelihood ratio of individual
effort, through log-supermodularity of $f_{X}$, while raising the market's
posterior assessment of talent. Part (iii) helps isolate this tension and is
not required for all results. Monotonicity is understood in the weak sense
throughout.

Assumption~\hyperlink{assm:1}{1} holds if $f_{\eta }$ is log-concave and
log-supermodular and the components of $\varepsilon $ are independent with
log-concave densities.\footnote{%
By Pr\'{e}kopa's theorem, convolution preserves log-concavity. With
independent log-concave noise, the $\mathrm{TP}_{2}$ composition and
marginalization results of Karlin and Rinott (\hyperlink{cite.KarlinRinott1980}%
{\textcolor{linkblue}{1980}}) imply log-supermodularity of $f_{X}$.} Under
Gaussianity, these sufficient conditions reduce to nonpositive off-diagonal
entries of $\Sigma _{\eta }^{-1}$ and diagonal $\Sigma _{\varepsilon }$.
Correlated noise is allowed, provided parts (ii) and (iii) hold, as
Proposition~\ref{prop:team}(ii) demonstrates.

Let $\sigma (Y)$ denote the information set generated by $Y$. The
performance rating for agent $i$, denoted by $r_{i}(Y)$, is any
square-integrable (i.e., finite-variance) statistic of $Y$: $r_{i}\in
L^{2}(\sigma (Y))\equiv L^{2}(Y)$. Examples include full disclosure, $%
r_{i}(Y)=Y$; team output, $r_{i}(Y)=\sum_{j}Y_{j}$; rank order, $r_{i}(Y)=%
\mathbf{1}\left\{ Y_{i}>\max_{j\neq i}Y_{j}\right\} $; and other nonlinear
and possibly vector-valued transformations of $Y$. I call the vector rule $%
r(Y)=(r_{1}(Y),\ldots ,r_{n}(Y))$ the \textbf{rating system}.\footnote{%
Ratings are deterministic functions of $Y$. In the Gaussian environment,
auxiliary randomization cannot raise effort: the optimal ratings in
Propositions~\ref{prop:relaxed}(ii) and \ref{prop:gauss} attain the values
of relaxed programs that allow it.}

The principal publicly commits to the rating system $r$ at the time of
hiring agents.\footnote{%
As previously noted, institutional guidelines, such as HR rating policies,
can ensure commitment. For instance, U.S. federal agencies' rating systems
are codified in \href{https://www.ecfr.gov/current/title-5/chapter-I/subchapter-B/part-430}%
{5 CFR Part~430}, recently amended by OPM (\hyperlink{cite.OPM2026}{%
\textcolor{linkblue}{2026}}), and professional bodies such as \href{https://www.shrm.org/topics-tools/topics/employee-performance-and-discipline}%
{SHRM} publish rating-policy guidance for private employers.} I initially
assume segmented labor markets: the market for agent $i$ observes only the
realized rating $r_{i}(Y)$; hence, ratings are \textbf{confidential}. This
fits settings where agents target different industries, geographies, or
internal divisions. Section~\ref{sec:public} considers \textbf{public}
ratings under which each agent's market observes the entire rating profile $%
r(Y)$.

Given the rating $r_{i}$ and conjectured effort profile $a^{\ast }$, the
market (inside or outside the organization) offers a wage equal to the
agent's expected talent: 
\begin{equation}
w_{i}(Y\text{ }|\text{ }a^{\ast })\equiv \mathbb{E}[\eta _{i}\text{ }|\text{ 
}r_{i}(Y),a^{\ast }]\text{.}  \label{market_wage}
\end{equation}%
In the two-period career-concerns interpretation, this wage equals expected
second-period output (since the last-period effort is zero in equilibrium,
and a new independent noise term is realized). I suppress these details to
save on notation.

Conjecturing the market wage and the peers' effort profile $a_{-i}^{\ast }$,
agent $i$ chooses effort to maximize net expected payoff: 
\begin{equation}
\max_{a_{i}}\mathbb{E}\left[ w_{i}(Y\text{ }|\text{ }a^{\ast })\text{ }|%
\text{ }a_{i},a_{-i}^{\ast }\right] -c_{i}(a_{i})\text{.}  \label{u_i_max}
\end{equation}%
The profile $a^{\ast }$ constitutes a (Bayes--Nash) equilibrium if it solves
(\ref{u_i_max}) for all $i$.

The principal earns no future profit because, in a competitive market,
workers are paid their expected outputs. She therefore chooses $r(Y)$ to
maximize expected first-period output (net of zero base wages), $\mathbb{E}%
\left[ \sum_{i}Y_{i}\text{ }|\text{ }a^{\ast }\right] $. Under (\ref%
{Output_i}), this is equivalent to maximizing total effort:\footnote{%
The principal's potential altruistic concern for agents' reputations would
not alter (P) because, by iterated expectations, ex ante expected talent is
invariant to the rating rule: $\mathbb{E}\left[ \mathbb{E}[\eta _{i}\text{ }|%
\text{ }r_{i},a]\text{ }|\text{ }a\right] =\mathbb{E}\left[ \eta _{i}\right] 
$ for all $r_{i}$ and $a$.} 
\begin{equation}
\max_{r(Y)}\text{ }\sum_{i}a_{i}^{\ast }\text{ \ \ s.t. \ }a^{\ast }\text{
is an equilibrium.}  \tag{P}
\end{equation}

The timing is as follows: the principal publicly commits to the rating
system $r(Y)$; agents simultaneously choose efforts $a$; uncertainty $(\eta
,\varepsilon )$ is realized and outputs $Y$ are generated; and the market
for agent $i$ observes $r_{i}(Y)$ and offers $w_{i}$.

\section{Characterization}

\label{sec:char}

Recall that $Y=X+a$, where the base output vector $X$ has density $f_{X}$.
Since effort enters as a location shift, the Fisher score (or the local
log-likelihood ratio) with respect to ``
parameter''\ $a_{i}$ is 
\begin{equation}
\frac{\partial \ln f_{X}(y-a)}{\partial a_{i}}=-\frac{\partial \ln f_{X}(x)}{%
\partial x_{i}}\equiv s_{i}(x).  \label{score_i_location}
\end{equation}%
The score $s_{i}(x)$ measures how favorable the observed output $y$ is for
inferring effort $a_{i}$: a positive score implies that $y$ points toward a
marginally higher $a_{i}$, while a negative score implies the opposite. As
usual, $s_{i}(X)$ has mean zero, and its variance is the Fisher information
about $a_{i}$.

Using (\ref{market_wage}), rewrite the wage schedule over base outputs as: 
\begin{equation}
w_{i}^{\ast }(X)\equiv \mathbb{E}[\eta _{i}\mid r_{i}(X+a^{\ast })]\text{.}
\label{wi_star}
\end{equation}%
Throughout, I focus on interior equilibria, where all agents exert positive
effort. The first result provides the equilibrium condition.

\begin{lemma}
\label{lem:FOC}\textup{(FOC)} Fix a rating system $r$. Any interior
equilibrium effort profile $a^{\ast }$ satisfies 
\begin{equation}
a_{i}^{\ast }=\gamma _{i}\operatorname{Cov}(w_{i}^{\ast }(X),s_{i}(X))\text{ \ \ }%
\forall i\text{.}  \label{FOC_Cov_i}
\end{equation}
\end{lemma}

The marginal benefit of effort is governed by the covariance between the
market's wage and the effort score: incentives are strong when the wage is
high exactly in those states where the output data are most indicative of
high effort. This is a familiar incentive constraint in agency problems,
except that, with career concerns, the market wage is the posterior belief
rather than a contractible transfer.\footnote{%
In the single-agent case, (\ref{FOC_Cov_i}) reduces to the first-order
condition (eq. (2.2)) in Dewatripont et al. (\hyperlink{cite.DewatripontEtAl1999a}%
{\textcolor{linkblue}{1999a}}).}

From (\ref{FOC_Cov_i}), the rating enters incentives only through the wage
it induces; the score $s_{i}$ is a primitive of the output technology.%
\footnote{%
Writing a deviation as $a_{i}=a_{i}^{\ast }+\delta $, the agent's payoff has
second derivative $\frac{d^{2}}{d\delta ^{2}}\mathbb{E}[w_{i}^{\ast
}(X+\delta e_{i})]-1/\gamma _{i}$. Thus, the FOC is globally sufficient if
the first term is everywhere below $1/\gamma _{i}$. This holds for
sufficiently small $\gamma _{i}$ under bounded curvature and, for the affine
Gaussian posterior wages characterized below, for every $\gamma _{i}$.}
Ratings are thus identified up to the induced posterior mean.

\begin{definition}
\label{def:equiv}For agent $i$, two rating rules are incentive-equivalent if
they induce the same posterior belief $w_{i}^{\ast }(X)$ almost surely.
\end{definition}

The next result simplifies the search for optimal ratings.

\begin{lemma}
\label{lem:scalar}\textup{(Scalar ratings)} Fix a rating rule $r_{i}$ and
equilibrium effort profile $a^{\ast }$. Then, the induced posterior $%
w_{i}^{\ast }(X)$ is an incentive-equivalent scalar rating, where $%
X=Y-a^{\ast }$. Moreover, it is Bayes plausible:%
\begin{equation}
\mathbb{E}[\eta _{i}\mid w_{i}^{\ast }(X)]=w_{i}^{\ast }(X).
\label{Bayes_Plausible}
\end{equation}
\end{lemma}

By definition of conditional expectation, $w_{i}^{\ast }(X)$ is the market's
best estimate of agent $i$'s talent given the rating $r_{i}$ and the
conjectured effort $a^{\ast }$. Directly reporting this estimate cannot
alter the market's posterior or the agent's incentives. In light of Lemma~%
\ref{lem:scalar}, I focus on characterizing $w_{i}^{\ast }(X)$.

Let 
\begin{equation*}
w_{i}^{f}(X)\equiv \mathbb{E}[\eta _{i}\mid X]
\end{equation*}%
denote the full-disclosure wage. By iterated expectations, (\ref%
{Bayes_Plausible}) is equivalent to: 
\begin{equation}
\mathbb{E}[w_{i}^{f}\mid w_{i}^{\ast }]=w_{i}^{\ast }.
\label{Bayes_Plausible_2}
\end{equation}%
The market's estimate of agent $i$'s talent based on the rating is
equivalent to inferring his full-disclosure wage. Because it suppresses $%
\eta _{i}$, (\ref{Bayes_Plausible_2}) allows writing the rating rule purely
as a function of output data. Also, since $\eta _{i}$ is mean-centered, so
are $w_{i}^{f}$ and $w_{i}^{\ast }$.

Combining Lemmas~\ref{lem:FOC} and \ref{lem:scalar}, program (P) reduces to
choosing the posterior wage vector $w^{\ast }$ solving 
\begin{equation}
\max_{w^{\ast }\in (L^{2}(X))^{n}}\sum_{i}\gamma _{i}\,\operatorname{Cov}%
(w_{i}^{\ast },s_{i}(X))\text{ \ s.t. \ }\mathbb{E}[w_{i}^{f}(X)\mid
w_{i}^{\ast }]=w_{i}^{\ast }\text{ \ }\forall i\text{.}  \tag{P1}
\end{equation}%
Although $w^{\ast }$ depends on $a^{\ast }$, its optimal choice does not:
under the additive-effort technology, expectations in (P1) are taken with
respect to the distribution of base output $X$. I therefore drop the
asterisks in the design problem. Moreover, (P1) is separable across agents;
for each $i$, 
\begin{equation}
\max_{w_{i}\in L^{2}(X)}\,\operatorname{Cov}(w_{i},s_{i}(X))\text{ \ s.t. \ }\mathbb{%
E}[w_{i}^{f}(X)\mid w_{i}]=w_{i}\text{.}  \tag{P2}
\end{equation}%
(P2) constitutes a Bayesian persuasion problem (Kamenica and Gentzkow 
\hyperlink{cite.KamenicaGentzkow2011}{\textcolor{linkblue}{2011}}) with a
continuous state space, whose constraint restricts the feasible posterior
means to the mean-preserving contractions of the \textit{effective} state $%
w_{i}^{f}(X)$. Importantly, (P2) depends only on the model primitives, $%
w_{i}^{f}(X)$ and $s_{i}(X)$. Let $w_{i}^{o}(X)$ solve (P2). By Lemma~\ref%
{lem:FOC}, equilibrium effort is given by 
\begin{equation*}
a_{i}^{\ast }=\gamma _{i}\operatorname{Cov}(w_{i}^{o}(X),s_{i}(X)).
\end{equation*}%
Accordingly, 
\begin{equation*}
r_{i}^{o}(Y)=w_{i}^{o}(Y-a^{\ast })
\end{equation*}%
is an optimal rating for agent $i$.

(P2) is not a standard persuasion problem in the posterior mean: since the
objective can be written as $\mathbb{E}[w_{i}\,\mathbb{E}[s_{i}\mid w_{i}]]$%
, each rating realization is valued by the product of the market's talent
estimate and the expected effort score.

The role of the Bayes-plausibility constraint in (P1) is also clear. Without
it, the principal could elicit arbitrarily large effort by setting the wage
(or equivalently, the rating) $w_{i}=ks_{i}$, which yields marginal benefits
proportional to the Fisher information, $k\operatorname{Var}(s_{i})$. But such a
rule is typically not a valid posterior under any rating rule.

Nevertheless, this suggests that an optimal rating would align the market
wage with the effort score as closely as Bayes plausibility permits. Under
Assumption~\hyperlink{assm:1}{1}, peer performance creates a tension between
the market wage and the effort score. The following assumption ensures that
this tension is nondegenerate.\medskip

\noindent \textbf{\hypertarget{assm:2}{Assumption 2.}} \textup{%
(Nondegenerate peer effect)} \textit{For each }$i$\textit{, both }$%
w_{i}^{f}(X)$\textit{\ and }$s_{i}(X)$\textit{\ are nonconstant in }$X_{j}$%
\textit{\ for all }$j\neq i$.

Assumptions~\hyperlink{assm:1}{1} and \hyperlink{assm:2}{2} are maintained
throughout unless stated otherwise.

\begin{lemma}
\label{lem:peer}\textup{(Peer tension)} The full-disclosure posterior $%
w_{i}^{f}(X)$ is increasing in $X$. The effort score $s_{i}(X)$ is
increasing in $X_{i}$ and decreasing in $X_{-i}$. Moreover, 
\begin{equation*}
\operatorname{Cov}(w_{i}^{f}(X),s_{i}(X)\text{ }|\text{ }X_{i})<0\text{ \ \ a.s.}
\end{equation*}
\end{lemma}

By Assumption~\hyperlink{assm:1}{1}(iii), under full disclosure, higher peer
performance is good news about agent $i$'s talent, raising the posterior
wage. At the same time, by Assumption~\hyperlink{assm:1}{1}(ii), it
partially explains $i$'s performance, lowering the effort score. Peer
performance thus moves the posterior wage and score in opposite directions,
reducing their covariance and weakening incentives. By contrast, own output
raises both. The following result is immediate.

\begin{proposition}
\label{prop:fulldisc}Full disclosure is not optimal: $r_{i}(Y)\neq Y$ for
all $i$.
\end{proposition}

The result follows because own-output disclosure, $r_{i}(Y)=Y_{i}$, elicits
strictly greater effort than full output disclosure: suppressing peer output
eliminates the negative covariance effect from Lemma~\ref{lem:peer}.
However, own-output disclosure is generically suboptimal, too, as
Proposition~\ref{prop:team}(i) demonstrates. It is optimal in the
single-agent benchmark (Remark~\hyperlink{rem:A1}{A1}), highlighting the
role of informational spillovers in multi-agent rating design.

To make further progress on (P2), note that its Bayes-plausibility
constraint can be written as an ``
all-tests''\ rationality check for the market: 
\begin{equation}
\mathbb{E}[(w_{i}^{f}(X)-w_{i})g(w_{i})]=0\text{, \ }\forall g\in
L^{2}(w_{i}).  \label{Ortho_1}
\end{equation}%
After observing $w_{i}$, the market's forecast error $w_{i}^{f}(X)-w_{i}$
must be orthogonal to \textit{every} square-integrable function $g(w_{i})$,
since conditional expectation is an orthogonal projection. That is, no
statistic of the disclosed rating can contain residual information about
full-information beliefs. Viewed this way, (P2) has infinitely many
rationality constraints.

As a first-order approximation, I consider a relaxed problem in which the
market only performs linear regressions of $w_{i}^{f}(X)$ based on $w_{i}$,
i.e., it only tests against $g(w_{i})=bw_{i}$. Since $\mathbb{E}[s_{i}(X)]=0$%
, the relaxed program becomes: 
\begin{equation}
\max_{w_{i}\in L^{2}(X)}\,\mathbb{E}[w_{i}s_{i}(X)]\text{ \ s.t. \ }\mathbb{E%
}[(w_{i}^{f}(X)-w_{i})w_{i}]=0\text{.}  \tag{P2-R}
\end{equation}%
Pointwise optimization results in an explicit solution.

\begin{proposition}
\label{prop:relaxed}\textup{(Relaxed confidential design)} (P2-R) has the
unique solution 
\begin{equation*}
w_{i}^{o,\mathrm{lin}}(X)=\frac{1}{2}\left( w_{i}^{f}(X)+\kappa
_{i}s_{i}(X)\right) \text{, \ \ }\kappa _{i}\equiv \lbrack \operatorname{Var}%
(w_{i}^{f})/\operatorname{Var}(s_{i})]^{1/2}\text{,}
\end{equation*}%
which induces effort $a_{i}^{\ast ,\mathrm{lin}}=\frac{\gamma _{i}}{2}\left( 
\operatorname{Cov}(w_{i}^{f},s_{i})+\sqrt{\operatorname{Var}(w_{i}^{f})\operatorname{Var}(s_{i})}%
\right) $. Let 
\begin{equation*}
\widehat{w}_{i}\equiv \mathbb{E}[w_{i}^{f}\mid w_{i}^{o,\mathrm{lin}}]\text{%
, \ \ }\Delta _{i}\equiv \frac{\gamma _{i}}{\kappa _{i}}\operatorname{Var}(w_{i}^{o,%
\mathrm{lin}}-\widehat{w}_{i})\text{, \ \ }\widehat{a}_{i}\equiv a_{i}^{\ast
,\mathrm{lin}}-\Delta _{i}\text{.}
\end{equation*}%
Then:

\begin{enumerate}
\item[(i)] Disclosing $w_{i}^{o,\mathrm{lin}}$ as the rating induces effort $%
\widehat{a}_{i}$, with 
\begin{equation*}
a_{i}^{f}\leq \widehat{a}_{i}\leq a_{i}^{\ast }\leq a_{i}^{\ast ,\mathrm{lin}%
}\text{,}
\end{equation*}%
where $a_{i}^{f}$ is effort under full disclosure.

\item[(ii)] If $(\eta ,\varepsilon )$ is jointly normal, the relaxation is
without loss: $w_{i}^{o,\mathrm{lin}}=w_{i}^{o}$ and $a_{i}^{\ast ,\mathrm{%
lin}}=a_{i}^{\ast }$.
\end{enumerate}
\end{proposition}

Proposition~\ref{prop:relaxed} characterizes the optimal wage when the
market infers talent from the rating only linearly. The wage aggregates the
full-disclosure posterior $w_{i}^{f}$ and the effort score $s_{i}$,
weighting them so that the two terms have equal variance. Because (P2-R)
relaxes the orthogonality requirements in (\ref{Ortho_1}), $a_{i}^{\ast ,%
\mathrm{lin}}$ bounds attainable effort from above. In general, the fewer
ways the market uses the rating to infer talent, the more closely the
principal can align the rating with the effort score, eliciting greater
effort.\footnote{%
One can tighten the relaxation by imposing further orthogonality conditions,
for example, $\mathbb{E}[(w_{i}^{f}(X)-w_{i})w_{i}^{k}]=0$, $k=2,3,\ldots$.}

Part (i) also establishes a computable lower bound on $a_{i}^{\ast }$ that
outperforms full disclosure: the principal discloses $w_{i}^{o,\mathrm{lin}}$%
, and the market rationally pays $\widehat{w}_{i}$, inducing effort $%
\widehat{a}_{i}$. The effort loss under the relaxed rating is at most $%
\Delta _{i}$, which is proportional to the variance of the
Bayes-plausibility gap. The effort interval $[\widehat{a}_{i},a_{i}^{\ast ,%
\mathrm{lin}}]$ is readily computable even in non-Gaussian environments: $%
w_{i}^{o,\mathrm{lin}}$ requires only $w_{i}^{f}$, $s_{i}$, and their second
moments; $\widehat{w}_{i}$ is a regression on a scalar statistic; and $%
\Delta _{i}$ is a residual second moment. In Example S1 of the Supplemental
Appendix, the interval never exceeds $3.6$ percent of its upper bound, while
disclosing $w_{i}^{o,\mathrm{lin}}$ raises effort by at least $25$ percent
over full disclosure.

Part (ii) shows that under joint normality, the linear rationality
restriction is without loss: the relaxed solution is Bayes plausible and
thus feasible in (P2). Moreover, since both $w_{i}^{f}(X)$ and $s_{i}(X)$
are affine, so is $w_{i}^{o}(X)$, implying that one may restrict attention
to \textit{linear} ratings of observed output $Y$: $r_{i}^{o}(Y)=\alpha
^{\top }Y$, \ $\alpha \in 
\mathbb{R}
^{n}$.\footnote{%
H\"{o}rner and Lambert (\hyperlink{cite.HornerLambert2021}{%
\textcolor{linkblue}{2021}}) also obtain linearity, in a dynamic Gaussian
setting with a single agent.} Next, I fully characterize the optimal rating
in the Gaussian environment and examine how it responds to peer output.

\medskip

\noindent \textbf{Gaussian environment.} Suppose 
\begin{equation*}
\eta \sim N(0,\Sigma _{\eta })\text{,\quad }\varepsilon \sim N(0,\Sigma
_{\varepsilon })\text{,\quad with }\sigma _{i}^{2}=\operatorname{Var}(\eta _{i})%
\text{, }\nu _{i}^{2}=\operatorname{Var}(\varepsilon _{i})\text{.}
\end{equation*}%
Assumptions~\hyperlink{assm:1}{1} and \hyperlink{assm:2}{2} are satisfied
when talents are strictly affiliated---meaning $\Sigma _{\eta }^{-1}$ has
strictly negative off-diagonal entries---and noise correlations are
sufficiently weak. In Proposition~\ref{prop:team}(ii) below, this condition
requires talent correlation to exceed noise correlation.

By Proposition~\ref{prop:relaxed}(ii), agent $i$'s rating weights peer $j$'s
performance by 
\begin{equation}
\partial _{j}w_{i}^{o}=\tfrac{1}{2}\underset{(+)}{(\underbrace{\partial
_{j}w_{i}^{f}}}+\kappa _{i}\,\underset{(-)}{\underbrace{\partial _{j}s_{i}})}%
\text{.}  \label{peer_loading}
\end{equation}%
Under normality, $w_{i}^{f}$ and $s_{i}$ are linear in $X$ (see (\ref{wf_s_X}%
)), rendering each weight constant. By Lemma~\ref{lem:peer}, their signs are
ambiguous and may differ across peers. I therefore classify ratings by their
aggregate peer weight.

\begin{definition}
\label{def:team}Let $T_{i}\equiv \sum_{j\neq i}\partial _{j}w_{i}^{o}$.
Agent $i$'s rating is called \textbf{team-based} if $T_{i}>0$ and \textbf{%
relative-performance} if $T_{i}<0$.
\end{definition}

To glean insight into $T_{i}$, I consider two special cases. Under
independent noise, 
\begin{equation}
w_{i}^{f}(X)=X_{i}-\nu _{i}^{2}s_{i}(X)\text{,}  \label{wi_f_si_X}
\end{equation}%
so peer output enters (\ref{peer_loading}) only through $s_{i}$, and all
peer weights share a sign. The exchangeable environment then isolates the
effect of correlated noise relative to correlated talent.

\begin{proposition}
\label{prop:team}\textup{(Team-based vs. relative-performance)}

\begin{enumerate}
\item[(i)] \textup{(Independent noise)} Suppose $\Sigma _{\varepsilon }=%
\operatorname{diag}(\nu _{1}^{2},\ldots ,\nu _{n}^{2})$. Then, for every $j\neq i$, 
\begin{equation*}
\operatorname{sign}\left( \partial _{j}w_{i}^{o}(X)\right) =\operatorname{sign}(1-\sigma
_{i}^{2}/\nu _{i}^{2})\text{.}
\end{equation*}

\item[(ii)] \textup{(Correlated noise)} Let $\sigma _{i}^{2}=\sigma ^{2}$, $%
\nu _{i}^{2}=\nu ^{2}$, $\operatorname{Corr}(\eta _{i},\eta _{j})=\rho $, and $\operatorname{Corr}(\varepsilon _{i},\varepsilon _{j})=\tau $ for all $i\neq j$, with $%
0\leq \tau <\rho <1$. Then, for every $j\neq i$, 
\begin{equation*}
\operatorname{sign}\left( \partial _{j}w_{i}^{o}(X)\right) =\operatorname{sign}(\tau ^{\ast
}-\tau )\text{,\quad }\tau ^{\ast }\equiv \frac{\rho }{2}(1-\sigma ^{2}/\nu
^{2})\text{.}
\end{equation*}
\end{enumerate}
\end{proposition}

Under independent noise, peer output matters only as a signal of agent $i$'s
talent. When own output is weakly informative of talent ($\sigma
_{i}^{2}/\nu _{i}^{2}<1$), the optimal rating is team-based: it leverages
collective reputation concerns to strengthen incentives. When own output is
highly informative ($\sigma _{i}^{2}/\nu _{i}^{2}>1$), the optimal rating is
relative: it nets out peer output, which exposes $i$'s rating to exogenous
variation without adding much information. Thus, under independent noise,
whether an agent's rating is team-based or relative depends only on his
signal-to-noise ratio, though peer weights depend on the entire information
structure. Interestingly, both rating forms may coexist within the same
organization---for example, across cohorts with different talent
uncertainty, tasks with different measurability, or assignments with
different exposure to exogenous shocks.

Part (ii) shows that positive noise correlation, perhaps reflecting shared
working conditions, favors relative-performance ratings. If, all else equal,
optimal ratings are team-based under independent noise ($\sigma ^{2}/\nu
^{2}<1$), they turn relative for sufficiently strong noise correlation,
i.e., $\tau >\tau ^{\ast }$. The mechanism is the familiar filtering motive
of contracting models (e.g., Holmstr\"{o}m \hyperlink{cite.Holmstrom1982}{%
\textcolor{linkblue}{1982}}, Section 3).

Nonetheless, the contrast with full disclosure---often assumed in the
literature following Holmstr\"{o}m (\hyperlink{cite.Holmstrom1999}{%
\textcolor{linkblue}{1999}})---is instructive. Meyer and Vickers (\hyperlink{cite.MeyerVickers1997}%
{\textcolor{linkblue}{1997}}) show that, with two symmetric agents, the
market's wage loads negatively on peer output only when noise is more
correlated than talent, $\tau >\rho $; that is, only when peer output
becomes bad news about agent $i$'s talent, violating Assumption~\hyperlink{assm:1}%
{1}(iii). Their result obtains from part (ii) by focusing on $w_{i}^{f}$
alone. In contrast, confidential design is relative-performance whenever $%
\tau >\tau ^{\ast }$, and $\tau ^{\ast }\leq \rho /2$, at most half the
correlation their result requires. The reason is that under full disclosure
the peer weight is $\partial _{j}w_{i}^{f}$ alone, reflecting peer evidence
about talent; confidential design adds a second term, $\kappa _{i}\partial
_{j}s_{i}$, peer evidence about effort, which is negative even under
independent noise.

The last result in this section elaborates on the contrast with full
disclosure.

\begin{proposition}
\label{prop:peerinfo}\textup{(Peer informativeness and incentives)} Suppose $%
\Sigma _{\varepsilon }=\operatorname{diag}(\nu _{1}^{2},\ldots ,\nu _{n}^{2})$. As
peer output becomes more informative about talent $\eta _{i}$, in the sense
of a higher $\operatorname{Var}(\mathbb{E}[\eta _{i}\mid X_{-i}])$ holding $\sigma
_{i}^{2}$ and $\nu _{i}^{2}$ fixed, effort decreases under full disclosure
but increases under the optimal rating.
\end{proposition}

\noindent Under full disclosure, more informative peer output weakens
incentives because the market attributes more of the agent's performance to
talent than to effort.\footnote{%
Under equicorrelated talents, $\operatorname{Var}(\mathbb{E}[\eta _{i}\mid X_{-i}])$
increases with talent correlation $\rho $.} Under optimal design, by
contrast, the principal uses this additional information to construct a
Bayes-plausible posterior wage that tracks the effort score more closely. In
a sense, full disclosure overuses peer output for talent inference,
exacerbating reputational free-riding, whereas optimal design uses peer
information selectively to maximize effort. This incentive reversal cautions
against applying full-disclosure intuition to hiring and training policies
that shape employee skill correlations.

\begin{remark}
\label{rem:size}\textup{(Number of agents)} The incentive advantage of
optimal design over full disclosure grows with organization size. Adding an
agent with independent noise, holding fixed the incumbents' joint
distribution, cannot lower $\operatorname{Var}(\mathbb{E}[\eta _{i}\mid X_{-i}])$,
since more peer output only spreads the market's posterior mean. By
Proposition~\ref{prop:peerinfo}, effort thus falls under full disclosure and
rises under the optimal rating: the gap, zero for a single agent, widens
with every peer informative about $\eta _{i}$.
\end{remark}

To understand the role of confidentiality, I next explore public ratings.

\section{Public ratings}

\label{sec:public}

Thus far, I have assumed separate labor markets, allowing each agent's
rating to remain confidential to his market. In practice, confidentiality is
often imperfect: ratings enter promotion reviews, inform job references, or
circulate through professional networks. This section studies the polar case
of \textbf{public ratings}, where all markets observe the entire rating
profile, $r(Y)$.

I begin with general properties of optimal public disclosure and then
specialize to the Gaussian environment for a complete characterization.

Given a rating system $r$ and the induced effort profile $a^{\ast }$, the
market wage for agent $i$ becomes 
\begin{equation*}
w_{i}^{\ast }(X)=\mathbb{E}[\eta _{i}\mid r(X+a^{\ast })]\text{.}
\end{equation*}%
Analogous to Lemma~\ref{lem:scalar}, the principal can equivalently choose
the posterior wage vector $w^{\ast }(X)$ as the rating profile (Lemma~
\hyperlink{lem:A1}{A1}). Under public design, however, Bayes plausibility is
a joint condition: once the market observes the profile $w$, each component $%
w_{i}$ must equal its posterior mean. Accordingly, in an interior
equilibrium, (P1) can be written as%
\begin{equation}
\max_{w\in \left( L^{2}(X)\right) ^{n}}\sum_{i}\gamma _{i}\,\operatorname{Cov}%
(w_{i},s_{i}(X))\text{ \ s.t. \ }\mathbb{E}[w_{i}^{f}(X)\mid w]=w_{i}\text{
\ }\forall i\text{.}  \tag{$\mathcal{P}1$}
\end{equation}%
Unlike its confidential counterpart, ($\mathcal{P}1$) is nonseparable across
agents. The first result shows that this informational leakage restricts the
principal, depressing equilibrium effort for every agent. Throughout the
remaining analysis, superscripts $c$ and $p$ denote optimal confidential and
public outcomes, respectively.

\begin{proposition}
\label{prop:subset}\textup{(Public} $\subseteq$ \textup{confidential)} Every
publicly feasible posterior is confidentially feasible. Hence:

\begin{enumerate}
\item[(i)] Public ratings elicit no more effort than confidential ratings: $%
a_{i}^{p}\leq a_{i}^{c}$ \ $\forall i$.

\item[(ii)] The principal prefers confidential ratings, whereas each agent
prefers public ratings.
\end{enumerate}
\end{proposition}

To see the logic, notice that, as in (\ref{Ortho_1}), the Bayes-plausibility
constraints in ($\mathcal{P}1$) are equivalent to requiring that, for each $%
i $,%
\begin{equation*}
\mathbb{E}[(w_{i}^{f}(X)-w_{i})g(w)]=0\quad \forall g\in L^{2}(w)\text{.}
\end{equation*}%
Since $g$ depends on the full vector $w$, these orthogonality conditions
reflect reputational spillovers across agents, making public disclosure more
restrictive than confidential disclosure. Under confidentiality, the
estimation error $w_{i}^{f}(X)-w_{i}$ need only be orthogonal to functions
of $w_{i}$.

Since the principal is more constrained under public design, she cannot
elicit higher effort from any agent and is therefore no better off than
under the confidential design. Intuitively, information leakage forces the
principal to forgo alignment between an agent's posterior belief and effort
score, weakening incentives. Notably, this simple revealed-preference
argument does not rely on Assumption~\hyperlink{assm:1}{1}.

Agents, by contrast, are no worse off under public ratings: by iterated
expectations, 
\begin{equation*}
\mathbb{E}[\mathbb{E}[\eta _{i}\mid r,a]\text{ }|\text{ }a]=\mathbb{E}[\eta
_{i}]
\end{equation*}%
for any rating system $r$ and effort profile $a$. Because expected wages
remain unchanged across regimes, lower equilibrium effort implies a higher
payoff for each agent under public ratings.

As with Proposition~\ref{prop:relaxed}, I first employ Lagrangian methods to
solve a relaxed version of ($\mathcal{P}1$). Under normality, this
relaxation is, again, without loss.

\begin{proposition}
\label{prop:relaxedpublic}\textup{(Relaxed public design)} Consider the
relaxation of \textrm{(}$\mathcal{P}1$\textrm{)} in which Bayes plausibility
is replaced by 
\begin{equation*}
\mathbb{E}\!\left[ (w^{f}-w)w^{\top }\right] =\mathbf{0}_{n\times n}.
\end{equation*}%
Any regular Lagrangian solution admits the representation 
\begin{equation}
w^{p,\mathrm{lin}}(X)=\left( \Psi +\Psi ^{\top }\right) ^{-1}\left( \Psi
^{\top }w^{f}(X)+\Gamma s(X)\right) ,  \label{w_i_P_lin}
\end{equation}%
where $\Psi \in \mathbb{R}^{n\times n}$ is the multiplier matrix, and
regularity means $\Psi +\Psi ^{\top }\succ 0$.

If $(\eta ,\varepsilon )$ is jointly normal, then $w^{p,\mathrm{lin}}(X)$
satisfies Bayes plausibility. Hence, the relaxation is without loss: $w^{p,%
\mathrm{lin}}$ solves ($\mathcal{P}1$).
\end{proposition}

Unlike under confidential design, each agent's public wage depends on the
full-disclosure wages and effort scores of all agents. Equation (\ref%
{w_i_P_lin}), however, provides little insight into the ``
richness''\ of optimal disclosure or whether full disclosure
is ever optimal. I thus turn my attention to Gaussian environments that
enable a spectral analysis and a full characterization. Later, I discuss the
generality of this approach.\medskip

\noindent \textbf{Gaussian environment.} Write ($\mathcal{P}1$) in matrix
form: 
\begin{equation}
\max_{w\in \left( L^{2}(X)\right) ^{n}}\mathcal{E}\equiv \mathbb{E}[w^{\top
}\Gamma s(X)]\text{ \ s.t. \ }\mathbb{E}[w^{f}(X)\mid w]=w 
\tag{$\mathcal{P}2$}
\end{equation}%
where $\Gamma \equiv \mathrm{diag}(\gamma _{1},\ldots ,\gamma _{n})$. As
hinted at in (\ref{wi_f_si_X}), a key property of the Gaussian setting is
that the full-disclosure posterior and score functions are linear
reparameterizations of each other: writing $\Sigma _{X}\equiv \operatorname{Var}(X)$%
, 
\begin{equation}
w^{f}(X)=\Sigma _{\eta }\Sigma _{X}^{-1}X\text{ \ \ and \ \ }s(X)=\Sigma
_{\eta }^{-1}w^{f}(X)\text{.}  \label{wf_s_X}
\end{equation}%
Hence, for any publicly feasible posterior $w$, $\mathbb{E}[s(X)\text{ }|%
\text{ }w]=\Sigma _{\eta }^{-1}\mathbb{E}[w^{f}(X)\text{ }|\text{ }w]=\Sigma
_{\eta }^{-1}w$. Substituting this into the principal's objective and
applying iterated expectations yields $\mathcal{E}=\mathbb{E}[\mathbb{E}%
[w^{\top }\Gamma s(X)\text{ }|\text{ }w]]=\mathbb{E}[w^{\top }\Gamma \Sigma
_{\eta }^{-1}w]$, which is quadratic in $w$. Although the cost-weighted
talent-precision matrix $\Gamma \Sigma _{\eta }^{-1}$ need not be symmetric,
only its symmetric part matters for this quadratic form. To this end, define 
\begin{equation*}
B\equiv \frac{\Gamma \Sigma _{\eta }^{-1}+\Sigma _{\eta }^{-1}\Gamma }{2}.
\end{equation*}%
Total effort is then 
\begin{equation}
\mathcal{E}=\mathbb{E}[w^{\top }Bw]=\mathrm{tr}\left( B\operatorname{Var}(w)\right) 
\text{.}  \label{Public_Effort}
\end{equation}%
The principal's objective becomes how much posterior wage variation to
preserve and along which directions, with $B$ measuring its incentive value.
The remaining issue in ($\mathcal{P}2$) is to operationalize the
Bayes-plausibility constraint. The next lemma separates this issue into a 
\textit{whitening }(decorrelation) step and, under normality, a simple
covariance characterization.

\begin{lemma}
\label{lem:BP}\textup{(Restating Bayes plausibility)} Let $Z$ be a centered
random vector with covariance matrix $\Sigma _{Z}\succ 0$, and let $%
\widetilde{Z}\equiv \Sigma _{Z}^{-1/2}Z$, so that $\operatorname{Var}(\widetilde{Z}%
)=I_{n}$. Given a random vector $W$, set $U\equiv \Sigma _{Z}^{-1/2}W$. Then:

\begin{enumerate}
\item[(i)] $\mathbb{E}[Z\mid W]=W$ if and only if $\mathbb{E}[\widetilde{Z}%
\mid U]=U$.

\item[(ii)] If $\widetilde{Z}\sim N(0,I_{n})$, a symmetric matrix $V$
satisfies $V=\operatorname{Var}(U)$ for some $U$ with $\mathbb{E}[\widetilde{Z}\mid
U]=U$ if and only if $0\preceq V\preceq I_{n}$.
\end{enumerate}
\end{lemma}

Part (i) does not rely on normality: whitening is just a linear
transformation of the Bayes-plausibility constraint, placing the information
structure in an orthogonal coordinate system. Part (ii) exploits normality
but only for sufficiency. Bayes plausibility requires that posterior means
have lower variance than the full-information object: $0\preceq V\preceq
I_{n}$. Normality ensures sufficiency: any covariance matrix satisfying
these bounds can be generated by a Bayes-plausible posterior mean, if
necessary with an auxiliary randomization.

Applying Lemma~\ref{lem:BP} with $Z=w^{f}$ and $W=w$, so that $w=\Sigma
_{w^{f}}^{1/2}U$ with $\Sigma _{w^{f}}\equiv \operatorname{Var}(w^{f})$, and noting
that 
\begin{equation*}
\operatorname{tr}(B\operatorname{Var}(w))=\operatorname{tr}(M\operatorname{Var}(U)),\qquad M\equiv \Sigma
_{w^{f}}^{1/2}B\Sigma _{w^{f}}^{1/2},
\end{equation*}%
($\mathcal{P}2$) reduces to a semidefinite program: 
\begin{equation}
\max_{V}\text{ }\operatorname{tr}(MV)\qquad \text{s.t.}\qquad 0\preceq V\preceq
I_{n}.  \tag{$\mathcal{P}3$}
\end{equation}%
Thus, the public design simplifies to choosing a posterior covariance $V$,
normalized so that the full-disclosure wage has covariance $I_{n}$. The
constraint permits the public rating to retain, but not amplify,
full-disclosure variation in any direction. With an infinitesimal cost per
disclosed direction to break ties, Proposition~\ref{prop:gauss} shows that
exactly the directions with positive marginal incentive value are preserved.

\begin{proposition}
\label{prop:gauss}\textup{(Full Gaussian characterization)} Let $M=Q\Lambda
Q^{\top }$ be an orthogonal diagonalization, with $\Lambda =\operatorname{diag}%
(\lambda _{1},\ldots ,\lambda _{n})$ and $Q=[q_{1},\ldots ,q_{n}]$, and set $%
J_{+}\equiv \{\ell :\lambda _{\ell }>0\}$ and $Q_{+}\equiv \left[ q_{\ell }%
\right] _{\ell \in J_{+}}$. Then, under optimal public design,

\begin{enumerate}
\item[(i)] total effort is $\mathcal{E}^{p}=\sum_{\ell \in J_{+}}\lambda
_{\ell }$;

\item[(ii)] the rating discloses the positive-eigenvalue coordinates and
suppresses the rest, with standardized representation 
\begin{equation*}
r^{p}(Y)\equiv Q_{+}^{\top }\Sigma _{w^{f}}^{-1/2}w^{f}(Y-a^{p})\sim
N(0,I_{|J_{+}|});
\end{equation*}

\item[(iii)] the induced posterior wage vector $w^{p}(Y)=\Sigma
_{w^{f}}^{1/2}Q_{+}r^{p}(Y)$ is unique.
\end{enumerate}
\end{proposition}

The characterization follows from the spectral decomposition of the
symmetric matrix $M$: the eigenvectors $q_{\ell }$ are orthogonal directions
of whitened full-disclosure posterior variation, and the eigenvalues $%
\lambda _{\ell }$ measure the incentive value of preserving variation along
them. The optimal public design retains only directions with positive value.%
\footnote{%
Zero-eigenvalue directions are withheld because of the infinitesimal
disclosure cost; disclosing them would leave total effort unchanged.}

Abusing notation, the standardized rating $r^{p}$ in part (ii) is a common
public signal, not a vector of individual ratings; equivalently, the
principal may disclose the posterior vector $w^{p}(Y)$ in part (iii), which
has $|J_{+}|$ linearly independent components. The optimal rating is not
unique, but its dimension, $|J_{+}|$, is.\footnote{%
Using (\ref{wf_s_X}) and suppressing the constant shift from $a^{p}$, the
uncentered rating is 
\begin{equation*}
\widetilde{r}^{p}(Y)=Q_{+}^{\top }\left[ \left( \Sigma _{\eta }(\Sigma
_{\eta }+\Sigma _{\varepsilon })^{-1}\Sigma _{\eta }\right) ^{1/2}\Sigma
_{\eta }^{-1}\right] Y.
\end{equation*}%
When $M$ is nonsingular, $w^{p}$ is also a regular Lagrangian solution in
Proposition~\ref{prop:relaxedpublic}; see Remark~\hyperlink{rem:A2}{A2}.}
Since the induced posterior is unique, garbling $r^{p}$ with nondegenerate
noise strictly lowers total effort.\footnote{%
Lemma~\ref{lem:BP}(ii) uses auxiliary noise to make every $V$ Bayes
plausible, making ($\mathcal{P}3$) a relaxation. Proposition~\ref{prop:gauss}
attains its solution with a deterministic rating.}

Notably, the dimension of the optimal rating depends exclusively on the
cost-weighted talent-precision matrix $B$: since $M$ and $B$ are congruent,%
\footnote{%
Because $\Sigma _{w^{f}}\succ 0$, its square root $\Sigma _{w^{f}}^{1/2}$
exists and is invertible, establishing the matrix congruence $M\equiv \left(
\Sigma _{w^{f}}^{1/2}\right) ^{\top }B\Sigma _{w^{f}}^{1/2}$.} Sylvester's
law of inertia implies that they share the same number of positive,
negative, and zero eigenvalues. Noise affects the rate of market learning
and hence the magnitudes of the eigenvalues and the output statistics
disclosed, but not whether preserving a given reputational direction
strengthens or weakens incentives. That sign is governed by the interaction
between talent dependence and cost heterogeneity, summarized by $B$.

Part (ii) readily implies that full disclosure (of $Y$) is strictly optimal
if and only if $B$ is positive definite. A simple sufficient condition is
that cost heterogeneity is sufficiently small.

\begin{corollary}
\label{cor:fullpublic}\textup{(Full public disclosure)} If $\max_{i}|\gamma
_{i}-\overline{\gamma }|/\overline{\gamma }<\lambda _{\min }(\Sigma _{\eta
})/\lambda _{\max }(\Sigma _{\eta })$, where $\overline{\gamma }\equiv \frac{%
1}{n}\sum_{i}\gamma _{i}$, then $B\succ 0$, rendering full disclosure
strictly optimal.
\end{corollary}

When agents' cost parameters are sufficiently close, because their tasks are
similar or the principal values their outputs similarly, the principal's
incentive objective aligns with full learning by the market. Agents respond
to reputational gains at similar marginal rates, so every direction of
posterior variation raises total effort. The noise covariance $\Sigma
_{\varepsilon }$ affects exact posterior wages, but not the optimality of
disclosing all performance data. As cost heterogeneity vanishes, the talent
covariance matrix $\Sigma _{\eta }$ also becomes immaterial for full
disclosure: if $\gamma _{i}=\gamma $ for all $i$, then $B=\gamma \Sigma
_{\eta }^{-1}\succ 0$, regardless of $\Sigma _{\eta }$ (since $\Sigma _{\eta
}\succ 0$).

This logic does not extend to confidential design, where the principal
tailors each agent's posterior wage without information leakage: by
Proposition~\ref{prop:fulldisc}, own-output disclosure strictly dominates
full disclosure even with identical costs, whereas under public design
disclosing own outputs is simply full disclosure.

Corollary~\ref{cor:fullpublic} has three useful implications. First, when
the optimal public rating is full disclosure, it is more informative in the
Blackwell sense than the optimal confidential rating for each agent. Under
partial public disclosure, however, the two regimes are not
Blackwell-ordered: as the numerical example below demonstrates, confidential
ratings may be more informative about some agents.

Second, under full disclosure, public ratings are clearly team-based: by
Lemma~\ref{lem:peer} and Assumption~\hyperlink{assm:2}{2}, the
full-disclosure wage $w_{i}^{f}(X)$, being linear, is strictly increasing in
peer performance $X_{j}$. This contrasts with Proposition~\ref{prop:team}%
(i), where confidential ratings may be relative. The reason is joint Bayes
plausibility: public design gives the principal less freedom to align each
agent's posterior wage with that agent's effort score---the very source of
relative-performance evaluation identified in Lemma~\ref{lem:peer}.

Third, as noted above, cost heterogeneity is necessary for partial public
disclosure. In particular, with ex ante identical agents as in Meyer and
Vickers (\hyperlink{cite.MeyerVickers1997}{\textcolor{linkblue}{1997}}), the
assumption of full output disclosure is without loss under public design.
The contrast in Section~\ref{sec:char} thus stems from confidentiality.

Corollary~\ref{cor:partial} shows that, among ex ante equally talented
agents, partial public disclosure is optimal when talent correlation exceeds
a threshold that decreases with cost heterogeneity.

\begin{corollary}
\label{cor:partial}\textup{(Partial disclosure)} Let talents be
equicorrelated, $\Sigma _{\eta }=(1-\rho )I_{n}+\rho \,\mathbf{1}\mathbf{1}%
^{\top }$ with $0<\rho <1$, and let $H(\Gamma )\equiv \lbrack
(\sum_{i}\gamma _{i})(\sum_{i}\gamma _{i}^{-1})]^{1/2}\geq n$ index cost
heterogeneity. Then the optimal public rating is $(n-1)$-dimensional if and
only if 
\begin{equation*}
\rho \geq \rho ^{\ast }\equiv \frac{2}{H(\Gamma )-n+2}.
\end{equation*}%
Otherwise it is $n$-dimensional, that is, full disclosure.
\end{corollary}

Cost heterogeneity and talent correlation reinforce each other.\footnote{%
The index $H$ measures cost heterogeneity because any mean-preserving spread
of the cost vector $\gamma $ strictly increases it.} A larger $H(\Gamma )$
raises the value of concentrating incentives on low-cost agents, lowering
the correlation required for partial disclosure. A larger $\rho $ makes
performance variation reflect common rather than idiosyncratic talent, so
public ratings move posterior beliefs together and the principal can less
easily align them with individual effort scores. Whenever costs are
heterogeneous, so that $H(\Gamma )>n$, a high enough $\rho $ therefore
drives the marginal incentive value of one aggregate direction to zero or
below, and the optimal rating suppresses it.\footnote{%
Under equicorrelated talents, public ratings of dimension below $n-1$ are
suboptimal; for $n\geq 3$ a rating based on total output alone, $%
\sum_{i}Y_{i}$, therefore cannot be optimal. Unequal talent correlations
combined with cost heterogeneity can support one-dimensional ratings (see
Example S2 in the Supplemental Appendix), but these generically differ from
the unweighted sum.} With identical costs, $H(\Gamma )=n$ and full
disclosure remains optimal for every $\rho <1$, consistent with Corollary~%
\ref{cor:fullpublic}.

Table 1a illustrates Corollary~\ref{cor:partial} for $n=3$ and $(\gamma
_{1},\gamma _{2},\gamma _{3})=(1,1,9)$, where $H(\Gamma )\approx 4.82$ and $%
\rho ^{\ast }\approx 0.52$. At $\rho =0.4$ all three eigenvalues of $M$ are
positive and full disclosure is optimal; at $\rho =0.6$ and $0.9$ one is
negative, reducing the optimal rating to two orthogonal components. Assuming
further $\varepsilon \sim N(0,I_{3})$ to characterize them, the optimal
rating combines a team-based statistic weighted heavily toward agent 3 with
an orthogonal relative-performance statistic for the symmetric pair.

\begin{center}
\textbf{Table 1a. Optimal Public Rating}

\begin{tabular}{ccccccc}
\hline\hline
$\rho $ & $\lambda (M)$ & $\mathcal{E}^{p}$ & $r^{p}$ & $a_{1}^{p}$ & $%
a_{2}^{p}$ & $a_{3}^{p}$ \\ \hline
$0.4$ & $(4.50,0.38,0.23)$ & $5.11$ & $(Y_{1},Y_{2},Y_{3})$ & $0.46$ & $0.46$
& $4.18$ \\ 
$0.6$ & $(4.51,0.29,-0.18)$ & $4.79$ & $\left( 0.03(Y_{1}+Y_{2})+Y_{3},\
Y_{1}-Y_{2}\right) $ & $0.15$ & $0.15$ & $4.49$ \\ 
$0.9$ & $(4.52,0.09,-1.24)$ & $4.61$ & $\left( 0.04(Y_{1}+Y_{2})+Y_{3},\
Y_{1}-Y_{2}\right) $ & $0.06$ & $0.06$ & $4.48$ \\ \hline\hline
\end{tabular}

{\scriptsize 
\parbox{\hsize}{\noindent \textit{Note.} $\Sigma _{\eta }=(1-\rho
)I_{3}+\rho \mathbf{1}\mathbf{1}^{\top }$, $\Sigma _{\varepsilon }=I_{3}$,
and $\gamma =(1,1,9)$. The reported rating components are mutually
orthogonal but nonstandardized. At $\rho =0.4$, they fully disclose $Y$.
Values are rounded to two decimal places.}}
\end{center}

\noindent Consider $\rho =0.6$. The first rating component, $%
0.03(Y_{1}+Y_{2})+Y_{3}$, carries most of the incentive value ($\lambda
_{1}=4.51$ versus $\lambda _{2}=0.29$). It keeps agent 3's posterior highly
sensitive to $Y_{3}$, while largely muting spillovers from $Y_{1}$ and $%
Y_{2} $: 
\begin{equation*}
w_{3}^{p}(Y)=0.01\left( Y_{1}+Y_{2}\right) +0.50Y_{3}-2.24.
\end{equation*}%
Agent 3 is not evaluated against agents 1 and 2, avoiding a loss of
individual talent information that would otherwise undermine his career
incentives. The second component, $Y_{1}-Y_{2}$, preserves
relative-performance evaluation within the symmetric high-cost pair,
inducing the symmetric posterior wages:%
\begin{equation*}
w_{i}^{p}(Y)=0.15Y_{i}-0.13Y_{3-i}+0.31Y_{3}-1.40,\text{ \ \ }i=1,2\text{.}
\end{equation*}%
The optimal public rating thus combines team-based evaluation centered on
agent 3 with a relative-performance signal that differentiates agents 1 and
2.

To compare informativeness across regimes, Table 1b reports the resulting
posterior wage variances. Because $w_{i}^{p}$ and $w_{i}^{c}$ are normally
distributed with the same mean, $\mathbb{E}[\eta _{i}]$, a higher variance
is equivalent to a mean-preserving spread and hence to greater Blackwell
informativeness about $\eta _{i}$ (Ganuza and Penalva \hyperlink{cite.GanuzaPenalva2010}%
{\textcolor{linkblue}{2010}}). In this example, each agent's signal-to-noise
ratio equals one, so (\ref{Proof_Cor_1_1}) implies $w_{i}^{c}(Y)=0.5Y_{i}$
and $r_{i}^{c}(Y)=Y_{i}$ for all $i$.

\begin{center}
\textbf{Table 1b. Posterior Variances}

\begin{tabular}{ccccc}
\hline\hline
$\rho $ & $\operatorname{Var}(w_{1}^{p})$ & $\operatorname{Var}(w_{2}^{p})$ & $\operatorname{Var}%
(w_{3}^{p})$ & $\operatorname{Var}(w_{i}^{c})$ \\ \hline
$0.4$ & $0.54$ & $0.54$ & $0.54$ & $0.50$ \\ 
$0.6$ & $0.26$ & $0.26$ & $0.52$ & $0.50$ \\ 
$0.9$ & $0.45$ & $0.45$ & $0.53$ & $0.50$ \\ \hline\hline
\end{tabular}
\end{center}

At $\rho =0.4$, the public rating fully discloses $Y$ and thus Blackwell
dominates every confidential rating. Under partial public disclosure ($\rho
\in \{0.6,0.9\}$), no uniform Blackwell ranking holds: public ratings
generate less posterior-wage dispersion for agents 1 and 2 but more for
agent 3 to concentrate incentives. Greater rating transparency does not
imply more information about every agent.\footnote{%
Example S3 in the Supplemental Appendix gives an environment in which
confidential ratings Blackwell dominate public ratings for every agent.}

\medskip

\noindent \textbf{The role of normality.} Normality enters Proposition~\ref%
{prop:gauss} at two steps: writing the public-design objective as a
quadratic form, (\ref{Public_Effort}), and replacing Bayes plausibility with
a covariance constraint. The first requires only that the effort score be
linear in the full-disclosure posterior, $s(X)=\Sigma _{\eta }^{-1}w^{f}(X)$%
. By Tweedie's formula, this holds whenever talent is Gaussian and noise is
independent, whatever the noise distribution (Lemma~\hyperlink{lem:A2}{A2}%
(i)); conversely, it requires Gaussian talent (Lemma~\hyperlink{lem:A2}{A2}%
(ii)).

For the second step, Bayes plausibility makes the residual $\zeta \equiv
w^{f}-w$ uncorrelated with $w$, so total effort is%
\begin{equation*}
\mathcal{E}=\mathcal{E}^{f}-\operatorname{tr}\left( B\operatorname{Var}(\zeta )\right) \text{%
.}
\end{equation*}%
The principal's instrument is thus the posterior variation she destroys, and 
$B$ prices it: concealing a direction pays precisely when that price is
nonpositive, whatever the noise distribution.

\begin{proposition}
\label{prop:noise}\textup{(Arbitrary noise)} Suppose that talent $\eta \sim
N(0,\Sigma _{\eta })$ and noise $\varepsilon $ is independent of $\eta $,
with $\mathbb{E}\Vert \varepsilon \Vert <\infty $. Suppose further that $%
w^{f}$ is absolutely continuous. Then full disclosure is strictly optimal
under public design if and only if $B\succ 0$.
\end{proposition}

\noindent Gaussian noise is thus not required for full disclosure. What it
delivers is the characterization when $B\nsucc 0$, so that optimal
disclosure is partial. As with Proposition~\ref{prop:subset}, the result
does not rely on Assumption~\hyperlink{assm:1}{1}.

\section{Extension: effort spillovers}

\label{sec:spillovers}In many organizations, agents' efforts also affect
each other's measured performance---\hspace{0pt}positively, as when software
developers review one another's code, or negatively, as when medical
residents compete for limited operating-room time. So far, I have assumed
away effort spillovers to isolate reputational spillovers across agents.
This section introduces them. For tractability, I specify agent $i$'s output
as\footnote{%
This specification follows the R\&D literature (e.g., Kamien et al. 
\hyperlink{cite.KamienEtAl1992}{\textcolor{linkblue}{1992}}) and has been
applied to teams in Yildirim (\hyperlink{cite.Yildirim2023}{%
\textcolor{linkblue}{2023}}).}%
\begin{equation}
Y_{i}=\eta _{i}+\varepsilon _{i}+a_{i}+\frac{\beta }{n-1}\sum_{j\neq
i}a_{j},\qquad -1<\beta <1\text{.}  \label{Output_spillover}
\end{equation}%
Thus, $\beta $ is the effect of an agent's effort on his peers' combined
output. The bounds ensure that effort matters more for an agent's own output
than for his peers' combined output. Agent $i$'s effort score becomes%
\footnote{%
With output (\ref{Output_spillover}), $Y=X+G_{\beta }a$ for $G_{\beta
}\equiv \left( 1-\frac{\beta }{n-1}\right) I_{n}+\frac{\beta }{n-1}\mathbf{1}%
\mathbf{1}^{\top }$; hence, as in (\ref{score_i_location}), $\partial \ln
f_{X}(y-G_{\beta }a)/\partial a_{i}=\sum_{j}(G_{\beta })_{ji}s_{j}(x)$.}%
\begin{equation}
\widetilde{s}_{i}(X)\equiv s_{i}(X)+\frac{\beta }{n-1}\sum_{j\neq i}s_{j}(X),
\label{score_spillover}
\end{equation}%
since peer performance is now also evidence of his effort: favorable
evidence when $\beta >0$ and unfavorable when $\beta <0$. Because each
agent's effort raises expected total output by the same factor, $1+\beta $,
the principal still maximizes total effort, and (P2) and ($\mathcal{P}2$)
are unchanged except that $\widetilde{s}_{i}$ replaces $s_{i}$. Lemmas~\ref%
{lem:FOC}--\ref{lem:scalar} and Proposition~\ref{prop:relaxed} therefore
apply verbatim. The optimal rating now depends on $\beta $, and so does its
net peer weight, $T_{i}\equiv \sum_{j\neq i}\partial _{j}w_{i}^{o}$, whose
sign classifies the rating as team-based ($T_{i}>0$) or relative-performance
($T_{i}<0$). More explicitly, I write $T_{i}(\beta )$.

\begin{corollary}
\label{cor:spillconf}\textup{(Confidential design)} In the Gaussian
environment of Section~\ref{sec:char} with output (\ref{Output_spillover}), $%
T_{i}^{\prime }(\beta )>0$ for every $i$.
\end{corollary}

Because peer output is evidence of an agent's effort, favorable when $\beta
>0$ and unfavorable when $\beta <0$, the optimal rating rewards it more when 
$\beta >0$ and penalizes it more when $\beta <0$. Positive spillovers thus
push the confidential rating toward team-based evaluation and negative
spillovers toward relative-performance evaluation, a rationale distinct from
reputational spillovers.

Under public design, Corollary~\ref{cor:fullpublic} shows that, absent
effort spillovers, partial disclosure requires cost heterogeneity: with
identical costs, full disclosure is strictly optimal for \textit{any} talent
and noise covariances. Spillovers of either sign can call for partial
disclosure, in a way that parallels cost heterogeneity. Just as Corollary~%
\ref{cor:partial} indexes cost heterogeneity by $H(\Gamma )$, let 
\begin{equation*}
H(\Sigma _{\eta })\equiv \left[ (\mathbf{1}^{\top }\Sigma _{\eta }\mathbf{1}%
)(\mathbf{1}^{\top }\Sigma _{\eta }^{-1}\mathbf{1})\right] ^{1/2}
\end{equation*}%
index talent heterogeneity. Since $(\Sigma _{\eta }\mathbf{1})_{i}=\operatorname{Cov}%
(\eta _{i},\sum_{j}\eta _{j})$ is agent $i$'s aggregate-talent exposure, $%
H(\Sigma _{\eta })$ captures departures from identical exposures: by
Cauchy--Schwarz, $H(\Sigma _{\eta })\geq n$, with equality if and only if
exposures are equal across agents, i.e., $\Sigma _{\eta }\mathbf{1}\propto 
\mathbf{1}$.\footnote{%
Let $\Sigma _{\eta }=\sum_{k}\pi _{k}p_{k}p_{k}^{\top }$ be the spectral
decomposition of $\Sigma _{\eta }$ and $\theta _{k}$ the angle between $%
p_{k} $ and $\mathbf{1}$. Then $H(\Sigma _{\eta })/n=\left[ \left(
\sum_{k}\pi _{k}\cos ^{2}\theta _{k}\right) \left( \sum_{k}\pi _{k}^{-1}\cos
^{2}\theta _{k}\right) \right] ^{1/2}$, the square root of the ratio of the
arithmetic to the harmonic mean of the eigenvalues $\pi _{k} $, with weights 
$\cos ^{2}\theta _{k}$ summing to one. Since $\pi _{k}=\operatorname{Var}%
(p_{k}^{\top }\eta )$, the index is larger when the principal components
that make up aggregate talent have more dispersed variances.}

\begin{proposition}
\label{prop:spill}\textup{(Public design)} In the Gaussian environment of
Section~\ref{sec:char} with identical costs, $\Gamma =\gamma I_{n}$, and
output (\ref{Output_spillover}), full disclosure is strictly optimal if and
only if $\beta _{\ast }<\beta <\beta ^{\ast }$; otherwise, the optimal
public rating has dimension $|J_{+}|=n-1$, where 
\begin{equation*}
\beta _{\ast }\equiv -\frac{2(n-1)}{H(\Sigma _{\eta })+n-2}\in \lbrack -1,0)%
\text{,\qquad }\beta ^{\ast }\equiv \frac{2(n-1)}{H(\Sigma _{\eta })-n+2}\in
(0,n-1]\text{.}
\end{equation*}
\end{proposition}

Proposition~\ref{prop:spill} shows that moderate spillovers of either sign
leave full disclosure optimal, whereas, when aggregate-talent exposures
differ, sufficiently strong ones call for partial disclosure despite
identical effort costs. As with cost heterogeneity in Corollary~\ref%
{cor:partial}, public information cannot be targeted: strong spillovers make
peer output substantial evidence of each agent's effort, but the market
credits it according to agents' aggregate-talent exposures. When exposures
differ, the principal concentrates incentives on some agents, suppressing
information through which others would free ride on reputations their effort
did not build. Notice also that the confidential rating varies continuously
in $\beta $ (Corollary~\ref{cor:spillconf}), whereas the public rating is
invariant to $\beta $ between $\beta _{\ast }$ and $\beta ^{\ast }$, where
it equals the full-disclosure wage, and changes discontinuously at the
thresholds.

\section{Discussion: rating systems in practice}

\label{sec:discussion}

The analysis speaks to the recurring debate over stack ranking and forced
distributions in performance management (see Wijayanti et al. \hyperlink{cite.WijayantiEtAl2025}%
{\textcolor{linkblue}{2025}} for a review). While peer-benchmarking systems
can sharpen differentiation, curb rating inflation, and improve
accountability, they can also filter out information about shared cohort
quality, weakening the reputational incentives that ratings are intended to
provide.

Private-sector experience illustrates this tension. Adobe, Microsoft,
Deloitte, Accenture, and General Electric moved away from stack ranking due
to concerns about collaboration, incentives, and retaining top talent. At
Microsoft, Eichenwald (\hyperlink{cite.Eichenwald2012}{%
\textcolor{linkblue}{2012}}) reports that `` superstars did
everything they could to avoid working alongside other top-notch
developers,''\ fearing that strong peers would lower their
relative rankings. Adobe faced a related retention problem: under its old
system, `` the people we wanted to keep would
leave.''\ Its replacement, Check-in, was designed to promote
`` agility, speed, and increased
collaboration''\ (Cohan \hyperlink{cite.Cohan2013}{%
\textcolor{linkblue}{2013}}). By contrast, Google, Meta, and Amazon have
recently tightened relative evaluation curves to raise performance standards
and manage labor costs.\footnote{%
See Elias (\hyperlink{cite.Elias2022}{\textcolor{linkblue}{2022}}) on
Google's GRAD system; Rodriguez and Horwitz (\hyperlink{cite.RodriguezHorwitz2023}%
{\textcolor{linkblue}{2023}}) on Meta's tightened review curves; and Kim (
\hyperlink{cite.Kim2024}{\textcolor{linkblue}{2024}}) on Amazon's
`` unregretted attrition''\ quotas.}

The same debate appears in the public sector. Historically, the U.S. Office
of Personnel Management (OPM) prohibited forced distributions, mandating
assessments against absolute standards (Fedweek \hyperlink{cite.Fedweek2019}{%
\textcolor{linkblue}{2019}}).\footnote{%
The ban constrained how ratings were produced, not how they were used:
absolute-standard ratings were still compared across employees for retention
and promotion.} In 2025, OPM lifted this ban for the Senior Executive
Service, and in 2026 extended forced distributions to General Schedule and
prevailing-rate employees (OPM \hyperlink{cite.OPM2025}{%
\textcolor{linkblue}{2025}}, \hyperlink{cite.OPM2026}{%
\textcolor{linkblue}{2026}}).

The results do not imply that relative ranking is inherently flawed. Rather,
viewing performance evaluation solely as a tool against rating inflation
overlooks a critical incentive cost for career-concerned employees. Whether
relative or team-based ratings better serve incentives depends partly on how
agents' efforts interact. When effort helps peers, peer output becomes
evidence of an agent's effort and warrants more weight (Corollary~\ref%
{cor:spillconf}). Some firms that abandoned stack ranking have moved in this
direction. Microsoft's replacement system evaluates not only individual
impact but also how employees draw on others' input and ``
what [they] contribute to others' success''\ (Bishop 
\hyperlink{cite.Bishop2013}{\textcolor{linkblue}{2013}}), and Deloitte's
redesign has team leaders rate the statement `` I would
always want him or her on my team''\ (Buckingham and Goodall 
\hyperlink{cite.BuckinghamGoodall2015}{\textcolor{linkblue}{2015}}). When
effort instead crowds out peers, as in competition for scarce resources, the
same logic favors relative evaluation.

The regime comparison also parallels pay secrecy. Because wages are
incentive-equivalent to ratings (Lemma~\ref{lem:scalar}), confidential and
public ratings correspond to pay secrecy and transparency. Practice is
consistent with Proposition~\ref{prop:subset} on both sides. Roughly 60
percent of U.S. private-sector workers report that their employer bans or
discourages discussing pay (Rosenfeld \hyperlink{cite.Rosenfeld2017}{%
\textcolor{linkblue}{2017}}), while regulation has increasingly pushed
toward transparency, most recently the EU Pay Transparency Directive.%
\footnote{%
Directive (EU) 2023/970, OJ L 132, 17.5.2023 (transposition deadline June 7,
2026).} Existing explanations emphasize bargaining (Cullen and Pakzad-Hurson 
\hyperlink{cite.CullenPakzadHurson2023}{\textcolor{linkblue}{2023}}) or, as
in Halac et al. (\hyperlink{cite.HalacEtAl2021}{\textcolor{linkblue}{2021}}%
), the coordination value of concealing relative standing when efforts are
complementary. The career-concerns channel here is distinct: confidentiality
strengthens incentives by aligning posterior wages with effort scores, even
absent effort spillovers or strategic complementarities. In practice, these
channels often coexist because employees care about both current pay and
future career prospects.

The results also offer a rationale for team-based evaluation that requires
neither technological complementarities nor effort spillovers. When agents
share selection standards or training---as in medical residencies, doctoral
cohorts, or civil-service classes---strong peer performance signals a strong
talent pool, creating a collective reputation. Consistent with this channel,
MacLeod et al. (\hyperlink{cite.MacLeodEtAl2017}{\textcolor{linkblue}{2017}}%
) find that Colombian employers reward a college's reputation---its
graduates' mean admission scores---and that the return to reputation fell,
while the return to individual scores rose, once an exit exam supplied a
personal signal of skill. Thus, when individual output is noisy, positive
peer weight can strengthen career incentives through collective reputation.

\section{Conclusion}

\label{sec:conclusion}

This paper studies how an organization should rate multiple career-concerned
agents, and whether it should keep their ratings confidential or make them
public. Peer performance plays two opposing roles: it informs the market
about an agent's talent but also helps distinguish talent from effort.
Optimal ratings therefore need not disclose all available information and
may reward or penalize strong peers depending on the informational
environment. Publicity constrains this tradeoff and weakens incentives: the
principal prefers confidentiality, whereas agents prefer transparency.

Endogenizing peer composition and dynamic learning are natural next steps.
As talent uncertainty resolves over a career, the results suggest that
optimal confidential ratings may shift from relative-performance toward
team-based evaluation (Proposition~\ref{prop:team}(i)). When some of that
uncertainty concerns a common component such as the firm's hiring standards,
learning also reduces talent correlation; with heterogeneous costs, optimal
public ratings may then move from partial toward full disclosure (Corollary~%
\ref{cor:partial}).

\newpage

\appendix\setcounter{equation}{0}\renewcommand{\theequation}{A-%
\arabic{equation}}

\begin{description}
\item[{\protect\Large Appendix}] 
\end{description}

\begin{proof}[Proof of Lemma~\protect\ref{lem:FOC}]
In (\ref{u_i_max}), agent $i$ takes $w_{i}(Y$ $|$ $a^{\ast })$ as given, so
effort enters only through the density of $Y$. Differentiating under the
integral sign, which Assumption~\hyperlink{assm:1}{1} permits, 
\begin{equation*}
\frac{\partial }{\partial a_{i}}\mathbb{E}\left[ w_{i}(Y\text{ }|\text{ }%
a^{\ast })\text{ }|\text{ }a_{i},a_{-i}^{\ast }\right] =\mathbb{E}\left[
w_{i}(Y\text{ }|\text{ }a^{\ast })\frac{\partial }{\partial a_{i}}\ln
f_{X}(Y-\left( a_{i},a_{-i}^{\ast }\right) )\text{ }|\text{ }%
a_{i},a_{-i}^{\ast }\right] \text{.}
\end{equation*}%
At $a_{i}=a_{i}^{\ast }$, where $Y-a^{\ast }=X$, equations (\ref%
{score_i_location}) and (\ref{wi_star}) reduce the right-hand side to $%
\mathbb{E}\left[ w_{i}^{\ast }(X)s_{i}(X)\right] =\operatorname{Cov}(w_{i}^{\ast
}(X),s_{i}(X))$, since $\mathbb{E}\left[ s_{i}(X)\right] =0$; equating it to
the marginal cost $a_{i}^{\ast }/\gamma _{i}$ yields (\ref{FOC_Cov_i}).
\end{proof}

\begin{proof}[Proof of Lemma~\protect\ref{lem:scalar}]
Since $w_{i}^{\ast }(X)=\mathbb{E}[\eta _{i}\mid r_{i}(X+a^{\ast })]$ is
measurable with respect to $r_{i}$, iterated expectations yield 
\begin{equation*}
\mathbb{E}[\eta _{i}\mid w_{i}^{\ast }(X)]=\mathbb{E}\left[ \mathbb{E}[\eta
_{i}\mid r_{i}(X+a^{\ast })]\mid w_{i}^{\ast }(X)\right] =w_{i}^{\ast }(X).
\end{equation*}%
Hence, directly disclosing the scalar $w_{i}^{\ast }(X)$ induces the same
posterior wage and is incentive-equivalent to $r_{i}$.
\end{proof}

\begin{proof}[Proof of Lemma~\protect\ref{lem:peer}]
By (\ref{score_i_location}), log-concavity and log-supermodularity of $f_{X}$
imply 
\begin{equation*}
\partial _{i}s_{i}(x)=-\partial _{ii}^{2}\ln f_{X}(x)\geq 0\text{, \ \ \ }%
\partial _{j}s_{i}(x)=-\partial _{ij}^{2}\ln f_{X}(x)\leq 0\text{ \ (}j\neq i%
\text{).}
\end{equation*}%
By Assumption~\hyperlink{assm:1}{1}(iii), $w_{i}^{f}(X)=\mathbb{E}[\eta
_{i}\mid X]$ is increasing in every component of $X$.

Conditioning on $X_{i}$ preserves log-supermodularity, so $X_{-i}\mid X_{i}$
is affiliated. Since $w_{i}^{f}(X)$ and $-s_{i}(X)$ are both increasing in $%
X_{-i}$, affiliation implies 
\begin{equation*}
\operatorname{Cov}(w_{i}^{f}(X),s_{i}(X)\text{ }|\text{ }X_{i})\leq 0\text{.}
\end{equation*}%
The inequality is strict a.s. by the strictly positive density in Assumption~
\hyperlink{assm:1}{1}(i) and the nondegeneracy in Assumption~\hyperlink{assm:2}%
{2}.
\end{proof}

\begin{proof}[Proof of Proposition~\protect\ref{prop:fulldisc}]
Let $w_{i}^{\mathrm{own}}=\mathbb{E}[\eta _{i}\mid X_{i}]$. By the law of
total covariance,%
\begin{equation*}
\operatorname{Cov}(w_{i}^{f},s_{i})=\operatorname{Cov}\!\left( \mathbb{E}[w_{i}^{f}\mid
X_{i}],\mathbb{E}[s_{i}\mid X_{i}]\right) +\mathbb{E}\!\left[ \operatorname{Cov}%
(w_{i}^{f},s_{i}\mid X_{i})\right] \text{.}
\end{equation*}%
Since, by iterated expectations, $\mathbb{E}[w_{i}^{f}\mid X_{i}]=w_{i}^{%
\mathrm{own}}$, which is $X_{i}$-measurable, the first term on the right
equals $\operatorname{Cov}(w_{i}^{\mathrm{own}},s_{i})$. Since the second term is
strictly negative by Lemma~\ref{lem:peer}, we have 
\begin{equation*}
\operatorname{Cov}(w_{i}^{f},s_{i})<\operatorname{Cov}(w_{i}^{\mathrm{own}},s_{i})\text{.}
\end{equation*}%
Lemma~\ref{lem:FOC} then implies that the own-output rating, $r_{i}(Y)=Y_{i}$%
, elicits strictly greater effort from agent $i$ than the fully disclosing
rating, $r_{i}(Y)=Y $. Hence, full disclosure is not optimal.
\end{proof}

\textbf{\hypertarget{rem:A1}{Remark A1.}} Full disclosure is optimal for a 
\textit{single} agent. The same covariance decomposition applied to any
rating $r$, with $w_{r}\equiv \mathbb{E}[\eta \mid r(Y)]$, implies $\operatorname{Cov}(w^{f},s)\geq \operatorname{Cov}(w_{r},s)$, because $w^{f}$ and $s$ are both
increasing in $Y$ by Assumption~\hyperlink{assm:1}{1}.

\begin{proof}[Proof of Proposition~\protect\ref{prop:relaxed}]
Fix $i$ and let $\psi $ be the multiplier on the constraint in (P2-R), so
that the Lagrangian for any $X$ is: $w_{i}s_{i}+\psi (w_{i}^{f}-w_{i})w_{i}$%
. Pointwise maximization requires 
\begin{equation}
s_{i}+\psi (w_{i}^{f}-2w_{i})=0\text{.}  \label{Proof_Prop_2_1}
\end{equation}%
Since $\operatorname{Var}(s_{i})>0$, $\psi \neq 0$, revealing $w_{i}=\frac{1}{2}%
(w_{i}^{f}+s_{i}/\psi )$. Substituting into the orthogonality constraint
yields $\psi ^{2}=\operatorname{Var}(s_{i})/\operatorname{Var}(w_{i}^{f})$. Maximization
requires the positive root: 
\begin{equation*}
\psi ^{o,\mathrm{lin}}=\sqrt{\frac{\operatorname{Var}(s_{i})}{\operatorname{Var}(w_{i}^{f})}}%
.
\end{equation*}%
Hence, the Lagrangian is strictly concave with unique maximizer $w_{i}^{o,%
\mathrm{lin}}$, and Lemma~\ref{lem:FOC} delivers $a_{i}^{\ast ,\mathrm{lin}}$%
.

For part (i), recall that every Bayes-plausible $w_{i}$ satisfies the linear
orthogonality in (P2-R) 
\begin{equation}
\mathbb{E}[(w_{i}^{f}-w_{i})w_{i}]=0,  \label{Proof_Prop_2_2}
\end{equation}%
as does $w_{i}^{o,\mathrm{lin}}$ by construction. Subtracting (\ref%
{Proof_Prop_2_2}) from that for $w_{i}^{o,\mathrm{lin}}$ gives 
\begin{equation}
\mathbb{E}[(w_{i}^{o,\mathrm{lin}}-w_{i})w_{i}^{f}]=\mathbb{E}[(w_{i}^{o,%
\mathrm{lin}})^{2}]-\mathbb{E}[w_{i}^{2}].  \label{Proof_Prop_2_2_1}
\end{equation}%
Moreover, (\ref{Proof_Prop_2_1}) implies $s_{i}=\psi ^{o,\mathrm{lin}%
}(2w_{i}^{o,\mathrm{lin}}-w_{i}^{f})$. Multiplying by $w_{i}^{o,\mathrm{lin}%
}-w_{i}$, taking expectations, and using (\ref{Proof_Prop_2_2_1}) yield 
\begin{equation}
\mathbb{E}[w_{i}^{o,\mathrm{lin}}s_{i}]-\mathbb{E}[w_{i}s_{i}]=\psi ^{o,%
\mathrm{lin}}\mathbb{E}[(w_{i}^{o,\mathrm{lin}}-w_{i})^{2}].
\label{Proof_Prop_2_3}
\end{equation}%
Let $\widehat{w}_{i}\equiv \mathbb{E}[w_{i}^{f}\mid w_{i}^{o,\mathrm{lin}}]$%
. By iterated expectations, $\mathbb{E}[w_{i}^{f}\mid \widehat{w}_{i}]=%
\widehat{w}_{i}$, so $\widehat{w}_{i}$ is Bayes plausible, and (\ref%
{Proof_Prop_2_3}) evaluated at $\widehat{w}_{i}$ reads 
\begin{equation*}
\widehat{a}_{i}\equiv \gamma _{i}\mathbb{E}[\widehat{w}_{i}s_{i}]=a_{i}^{%
\ast ,\mathrm{lin}}-\Delta _{i},\qquad \Delta _{i}\equiv \gamma _{i}\psi ^{o,%
\mathrm{lin}}\operatorname{Var}(w_{i}^{o,\mathrm{lin}}-\widehat{w}_{i})\text{,}
\end{equation*}%
because $w_{i}^{o,\mathrm{lin}}-\widehat{w}_{i}$ is centered. Since $%
\widehat{w}_{i}$ is feasible for (P2), $a_{i}^{\ast }\geq \widehat{a}_{i}$,
while $a_{i}^{\ast }\leq a_{i}^{\ast ,\mathrm{lin}}$ because (P2-R) relaxes
(P2).

It remains to compare $\widehat{w}_{i}$ with full disclosure. Applying (\ref%
{Proof_Prop_2_3}) at $\widehat{w}_{i}$ and at $w_{i}^{f}$, 
\begin{eqnarray*}
\mathbb{E}[\widehat{w}_{i}s_{i}]-\mathbb{E}[w_{i}^{f}s_{i}] &=&\psi ^{o,%
\mathrm{lin}}\left( \operatorname{Var}(w_{i}^{o,\mathrm{lin}}-w_{i}^{f})-\operatorname{Var}%
(w_{i}^{o,\mathrm{lin}}-\widehat{w}_{i})\right) \\
&=&\psi ^{o,\mathrm{lin}}\mathbb{E}[\operatorname{Var}(w_{i}^{f}\mid w_{i}^{o,%
\mathrm{lin}})]\geq 0\text{,}
\end{eqnarray*}%
where the second equality applies the law of total variance to $w_{i}^{o,%
\mathrm{lin}}-w_{i}^{f}$ given that $w_{i}^{o,\mathrm{lin}}-\widehat{w}_{i}=%
\mathbb{E}[w_{i}^{o,\mathrm{lin}}-w_{i}^{f}\mid w_{i}^{o,\mathrm{lin}}]$.
Together, we find 
\begin{equation*}
a_{i}^{f}\leq \widehat{a}_{i}\leq a_{i}^{\ast }\leq a_{i}^{\ast ,\mathrm{lin}%
}\text{.}
\end{equation*}

For part (ii), joint normality of $(\eta ,\varepsilon )$ makes $X$ normal,
and $w^{f}$ and $s$ linear in $X$ (see (\ref{wf_s_X})). Therefore, $w_{i}^{o,%
\mathrm{lin}}$ is linear in $X$ and $(w_{i}^{f},w_{i}^{o,\mathrm{lin}})$ is
jointly normal. Since both variables are centered, the relaxed constraint (%
\ref{Proof_Prop_2_2}) implies $\operatorname{Cov}(w_{i}^{f},w_{i}^{o,\mathrm{lin}})=%
\operatorname{Var}(w_{i}^{o,\mathrm{lin}})$, and, in turn, $\mathbb{E}[w_{i}^{f}\mid
w_{i}^{o,\mathrm{lin}}]=w_{i}^{o,\mathrm{lin}}$. Thus $w_{i}^{o,\mathrm{lin}%
} $ is feasible for (P2), implying $w_{i}^{o,\mathrm{lin}}=w_{i}^{o}$ and $%
a_{i}^{\ast ,\mathrm{lin}}=a_{i}^{\ast }$.
\end{proof}

\begin{proof}[Proof of Proposition~\protect\ref{prop:team}]
Under normality, $w^{f}(X)=\Sigma _{\eta }\Sigma _{X}^{-1}X$ and $%
s(X)=\Sigma _{X}^{-1}X$. Thus, $\operatorname{Var}(s_{i})=1/\upsilon _{i}^{2}$,
where $\upsilon _{i}^{2}\equiv \operatorname{Var}(X_{i}\mid X_{-i})$. Moreover, $%
\operatorname{Cov}(X,s)=I_{n}$, and, by Stein's lemma, for any linear $g(X)=b^{\top
}X$, 
\begin{equation}
\mathbb{E}[\partial _{j}g]=b_{j}=\operatorname{Cov}(g,s_{j})\text{.}
\label{deriv_cov}
\end{equation}

(i) With independent noise, $\mathbb{E}[\varepsilon _{i}\mid X]=\nu
_{i}^{2}s_{i}(X)$, yielding $w_{i}^{f}=X_{i}-\nu _{i}^{2}s_{i}$. Hence 
\begin{equation}
\kappa _{i}^{2}=\nu _{i}^{4}+(\sigma _{i}^{2}-\nu _{i}^{2})\upsilon _{i}^{2}%
\text{,\qquad }w_{i}^{o}(X)=\frac{1}{2}\left[ X_{i}+(\kappa _{i}-\nu
_{i}^{2})s_{i}(X)\right] .  \label{Proof_Cor_1_1}
\end{equation}
Lemma~\ref{lem:FOC} then yields 
\begin{equation}
a_{i}^{o}=\frac{\gamma _{i}}{2}\left( \frac{\kappa _{i}+\sigma _{i}^{2}}{%
\kappa _{i}+\nu _{i}^{2}}\right) .  \label{Proof_Cor_1_2}
\end{equation}
For $j\neq i$, $\partial _{j}w_{i}^{o}=\frac{1}{2}(\kappa _{i}-\nu
_{i}^{2})\partial _{j}s_{i}$. By Lemma~\ref{lem:peer}, $\partial _{j}s_{i}<0$%
, while $\operatorname{sign}(\kappa _{i}-\nu _{i}^{2})=\operatorname{sign}(\sigma
_{i}^{2}-\nu _{i}^{2})$, proving part (i).

(ii) Let $\omega _{e}\equiv \sigma ^{2}(1-\rho )+\nu ^{2}(1-\tau )$ and $%
\omega _{m}\equiv \sigma ^{2}\rho +\nu ^{2}\tau $. Since $\Sigma _{X}=\omega
_{e}I_{n}+\omega _{m}\mathbf{1}\mathbf{1}^{\top }$, for $j\neq i$, 
\begin{equation*}
\partial _{j}s_{i}=-\frac{\omega _{m}}{\omega _{e}(\omega _{e}+n\omega _{m})}%
\text{,\qquad }\partial _{j}S=\frac{1}{\omega _{e}+n\omega _{m}},
\end{equation*}%
where $S\equiv \sum_{k}s_{k}$. Since $w_{i}^{f}=\sigma ^{2}(1-\rho
)s_{i}+\sigma ^{2}\rho S$, Proposition~\ref{prop:relaxed} implies 
\begin{equation*}
2\partial _{j}w_{i}^{o}=\frac{L-\kappa _{i}\omega _{m}}{\omega _{e}(\omega
_{e}+n\omega _{m})}\text{,\qquad }L\equiv \sigma ^{2}\nu ^{2}(\rho -\tau )>0.
\end{equation*}%
Moreover, 
\begin{equation*}
\kappa _{i}^{2}=\frac{\sigma ^{4}\left[ (1-\rho )^{2}(\omega
_{e}+(n-1)\omega _{m})+\rho (2-2\rho +n\rho )\omega _{e}\right] }{\omega
_{e}+(n-1)\omega _{m}},
\end{equation*}%
revealing 
\begin{equation}
L^{2}-\kappa _{i}^{2}\omega _{m}^{2}=\frac{\rho \sigma ^{4}\omega
_{e}(\omega _{e}+n\omega _{m})}{\omega _{e}+(n-1)\omega _{m}}\,\Xi \text{%
,\qquad }\Xi \equiv \rho (\nu ^{2}-\sigma ^{2})-2\nu ^{2}\tau .
\label{Xi_factor}
\end{equation}%
Since $L+\kappa _{i}\omega _{m}>0$, (\ref{Xi_factor}) implies $\operatorname{sign}%
(\partial _{j}w_{i}^{o})=\operatorname{sign}(\Xi )=\operatorname{sign}(\tau ^{\ast }-\tau )$%
, where $\tau ^{\ast }=\frac{\rho }{2}(1-\sigma ^{2}/\nu ^{2})$. This proves
part (ii).
\end{proof}

\begin{proof}[Proof of Proposition~\protect\ref{prop:peerinfo}]
Let $q_{i}\equiv \operatorname{Var}(\mathbb{E}[\eta _{i}\mid X_{-i}])$. Since $%
\varepsilon _{i}$ is independent of $(\eta _{i},X_{-i})$, joint normality
and the law of total variance yield 
\begin{equation*}
\upsilon _{i}^{2}=\sigma _{i}^{2}+\nu _{i}^{2}-q_{i}\text{.}
\end{equation*}%
Under full disclosure, 
\begin{equation*}
a_{i}^{f}=\gamma _{i}\left(1-\frac{\nu _{i}^{2}}{\upsilon _{i}^{2}}\right),
\end{equation*}
which decreases in $q_{i}$.

Under the optimal rating, (\ref{Proof_Cor_1_2}) and $\kappa _{i}^{2}=\nu
_{i}^{4}+(\sigma _{i}^{2}-\nu _{i}^{2})\upsilon _{i}^{2}$ imply 
\begin{equation*}
2\kappa _{i}\frac{\partial \kappa _{i}}{\partial q_{i}}=-(\sigma
_{i}^{2}-\nu _{i}^{2}).
\end{equation*}%
Hence, 
\begin{equation*}
\frac{\partial a_{i}^{o}}{\partial q_{i}}=\frac{\gamma _{i}(\sigma
_{i}^{2}-\nu _{i}^{2})^{2}}{4\kappa _{i}(\kappa _{i}+\nu _{i}^{2})^{2}}\geq
0,
\end{equation*}%
with strict inequality whenever $\sigma _{i}^{2}\neq \nu _{i}^{2}$.
\end{proof}

\begin{description}
\item \textbf{\hypertarget{lem:A1}{Lemma A1.}} \textit{Fix }$r$\textit{\ and
let }$w^{\ast }\equiv \mathbb{E}[\eta \mid r(Y),a^{\ast }]$. \textit{Then }$%
\mathbb{E}[w_{i}^{f}(X)\mid w^{\ast }]=w_{i}^{\ast }$\textit{\ }$\forall i$%
\textit{.}
\end{description}

\begin{proof}
Because $a^{\ast }$ is deterministic, $X=Y-a^{\ast }$ and $Y$ generate the
same information, so $w_{i}^{f}(X)=\mathbb{E}[\eta _{i}\mid Y,a^{\ast }]$.
Since $w^{\ast }=\mathbb{E}[\eta \mid r(Y),a^{\ast }]$ is measurable with
respect to $r(Y)$, iterated expectations yield 
\begin{equation*}
\mathbb{E}[w_{i}^{f}(X)\mid w^{\ast }]=\mathbb{E}[\eta _{i}\mid w^{\ast }]=%
\mathbb{E}[w_{i}^{\ast }\mid w^{\ast }]=w_{i}^{\ast }.
\end{equation*}
\end{proof}

\begin{proof}[Proof of Proposition~\protect\ref{prop:subset}]
Let $w$ be publicly feasible, so that $\mathbb{E}[w_{i}^{f}\mid w]=w_{i}$
for each $i$. Since $\sigma (w_{i})\subseteq \sigma (w)$, iterated
expectations give 
\begin{equation*}
\mathbb{E}[w_{i}^{f}\mid w_{i}]=\mathbb{E}\left[ \mathbb{E}[w_{i}^{f}\mid
w]\mid w_{i}\right] =\mathbb{E}[w_{i}\mid w_{i}]=w_{i}\text{,}
\end{equation*}%
so $w$ is confidentially feasible. Therefore, the optimal public posteriors $%
w^{p}$ are feasible in the confidential problem, and optimality of $%
w_{i}^{c} $ implies $\operatorname{Cov}(w_{i}^{p},s_{i}(X))\leq \operatorname{Cov}%
(w_{i}^{c},s_{i}(X))$ for each $i$. Then, by Lemma~\ref{lem:FOC}, $%
a_{i}^{p}\leq a_{i}^{c}$, which is part (i).

For part (ii), the principal prefers confidential ratings because effort is
higher for each agent. Each agent prefers public ratings because Bayes
plausibility makes expected wages identical, $\mathbb{E}[w_{i}^{p}]=\mathbb{E%
}[w_{i}^{f}]=\mathbb{E}[w_{i}^{c}]$, while effort and hence effort cost are
lower.
\end{proof}

\begin{proof}[Proof of Proposition~\protect\ref{prop:relaxedpublic}]
Let $\Psi \in \mathbb{R}^{n\times n}$ be the multiplier matrix on the
relaxed constraint and set $\Omega \equiv \Psi +\Psi ^{\top }$, so that
regularity is $\Omega \succ 0$. Using $\mathrm{tr}(\Psi ^{\top }ab^{\top
})=b^{\top }\Psi ^{\top }a$ and $w^{\top }\Psi ^{\top }w=\frac{1}{2}w^{\top
}\Omega w$, the Lagrangian is 
\begin{equation*}
\mathcal{L}(w;\Psi )=\mathbb{E}[w^{\top }\Gamma s]+\mathrm{tr}\left( \Psi
^{\top }\mathbb{E}[(w^{f}-w)w^{\top }]\right) =\mathbb{E}[w^{\top }\Gamma
s+w^{\top }\Psi ^{\top }w^{f}-\frac{1}{2}w^{\top }\Omega w]\text{.}
\end{equation*}%
Its pointwise Hessian is $-\Omega $, so a maximum requires $\Omega \succeq 0$%
, and the first-order condition $\Omega w^{p,\mathrm{lin}}=\Gamma s+\Psi
^{\top }w^{f}$ gives (\ref{w_i_P_lin}) under regularity.

The constraint also pins down $\Psi $. Let $K\equiv \operatorname{Cov}(\Gamma
s,w^{f})$. Substituting (\ref{w_i_P_lin}) into $\mathbb{E}[(w^{f}-w^{p,%
\mathrm{lin}})(w^{p,\mathrm{lin}})^{\top }]=0_{n\times n}$ and multiplying
through by $\Omega $ yields 
\begin{equation*}
\Omega \left( K^{\top }+\Sigma _{w^{f}}\Psi \right) =\operatorname{Var}(\Gamma
s)+K\Psi +\Psi ^{\top }K^{\top }+\Psi ^{\top }\Sigma _{w^{f}}\Psi \text{.}
\end{equation*}%
Expanding $\Omega =\Psi +\Psi ^{\top }$ on the left cancels the last two
terms on the right, leaving the Riccati equation 
\begin{equation}
\Psi \Sigma _{w^{f}}\Psi +\Psi K^{\top }-K\Psi -\operatorname{Var}(\Gamma
s)=0_{n\times n}\text{.}  \label{Proof_Prop_6_Riccati}
\end{equation}

Now let $(\eta ,\varepsilon )$ be jointly normal. Since $w^{f}$ and $s$ are
affine in $X$, so is $w^{p,\mathrm{lin}}$. Moreover, $u\equiv w^{f}-w^{p,%
\mathrm{lin}}$ is jointly normal with $w^{p,\mathrm{lin}}$. Then, the
relaxed constraint reads $\operatorname{Cov}(u,w^{p,\mathrm{lin}})=0$, implying $u$
and $w^{p,\mathrm{lin}}$ are independent and 
\begin{equation*}
\mathbb{E}\left[ w^{f}\mid w^{p,\mathrm{lin}}\right] =w^{p,\mathrm{lin}}+%
\mathbb{E}\left[ u\mid w^{p,\mathrm{lin}}\right] =w^{p,\mathrm{lin}}\text{.}
\end{equation*}%
Since $\Omega \succ 0$, the Lagrangian is strictly concave in $w$, so $w^{p,%
\mathrm{lin}}$ maximizes it and, satisfying the relaxed constraint, solves
the relaxed problem. Being Bayes plausible, it also solves ($\mathcal{P}1$).
\end{proof}

\begin{proof}[Proof of Lemma~\protect\ref{lem:BP}]
For part (i), since $\Sigma _{Z}^{1/2}$ is invertible, $\sigma (W)=\sigma
(U) $. Hence, 
\begin{equation*}
\mathbb{E}[Z\mid W]=\Sigma _{Z}^{1/2}\mathbb{E}[\widetilde{Z}\mid U],
\end{equation*}%
so $\mathbb{E}[Z\mid W]=W$ if and only if $\mathbb{E}[\widetilde{Z}\mid U]=U$%
.

For part (ii), if $\mathbb{E}[\widetilde{Z}\mid U]=U$, the law of total
variance implies 
\begin{equation*}
0\preceq V\equiv \operatorname{Var}(U)\preceq \operatorname{Var}(\widetilde{Z})=I_{n}.
\end{equation*}%
Conversely, let $0\preceq V\preceq I_{n}$, implying $V-V^{2}\succeq 0$. Set 
\begin{equation*}
U=V\widetilde{Z}+\xi \text{, \ \ }\xi \sim N(0,V-V^{2})\text{, \ \ }\xi
\perp \widetilde{Z}.
\end{equation*}%
Then, $\operatorname{Var}(U)=V$ and $\operatorname{Cov}(\widetilde{Z},U)=V$. Joint normality
implies 
\begin{equation*}
\mathbb{E}[\widetilde{Z}\mid U]=VV^{+}U=U,
\end{equation*}%
where $V^{+}$ denotes the Moore--Penrose pseudoinverse, which equals $V^{-1}$
when $V$ is nonsingular. The last equality follows because $VV^{+}$ is the
orthogonal projection onto the range of $V$, and $\operatorname{Var}(U)=V$ implies
that $U$ lies in that range almost surely.
\end{proof}

\begin{proof}[Proof of Proposition~\protect\ref{prop:gauss}]
Let $M=Q\Lambda Q^{\top }$ be the orthogonal diagonalization in the
statement. For feasible $V$, define $\widehat{V}\equiv Q^{\top }VQ$. Since $%
\widehat{V}$ is orthogonally similar to $V$, $0\preceq \widehat{V}\preceq
I_{n}$, so $0\leq \widehat{V}_{\ell \ell }\leq 1$. By the cyclic property of
the trace, 
\begin{equation*}
\operatorname{tr}(MV)=\operatorname{tr}(\Lambda \widehat{V}) =\sum_{\ell =1}^{n}\lambda
_{\ell }\widehat{V}_{\ell \ell }.
\end{equation*}%
Hence total effort is maximized by setting $\widehat{V}_{\ell \ell }=1$ when 
$\lambda _{\ell }>0$ and $\widehat{V}_{\ell \ell }=0$ when $\lambda _{\ell
}<0$, yielding 
\begin{equation*}
\mathcal{E}^{p}=\sum_{\ell \in J_{+}}\lambda _{\ell }.
\end{equation*}%
Under the infinitesimal disclosure cost, zero-eigenvalue directions are also
suppressed, selecting 
\begin{equation*}
\widehat{V}^{\ast }=\operatorname{diag}(1_{\{\lambda _{1}>0\}},\ldots , 1_{\{\lambda
_{n}>0\}}).
\end{equation*}%
This selection is unique: if $\widehat{V}_{\ell \ell }=0$, positive
semidefiniteness forces the $\ell$th row and column of $\widehat{V}$ to
vanish; if $\widehat{V}_{\ell \ell }=1$, the same argument applied to $I_{n}-%
\widehat{V}$ forces the remaining entries in that row and column to vanish.
Thus 
\begin{equation*}
V^{\ast }=Q\widehat{V}^{\ast }Q^{\top }=\sum_{\ell \in J_{+}}q_{\ell
}q_{\ell }^{\top }.
\end{equation*}

For part (ii), we apply Lemma~\ref{lem:BP} with $Z=w^{f}$ and $W=w$, so that 
$\widetilde{w}^{f}\equiv \Sigma _{w^{f}}^{-1/2}w^{f}$. Let $\mathcal{G}$ be
generated by 
\begin{equation*}
r_{\ell }^{p}(X)\equiv q_{\ell }^{\top }\widetilde{w}^{f}(X),\qquad \ell \in
J_{+}.
\end{equation*}%
Since $\widetilde{w}^{f}\sim N(0,I_{n})$ and the $q_{\ell }$'s are
orthonormal, these coordinates are independent $N(0,1)$. Hence 
\begin{equation*}
U^{\ast }\equiv \mathbb{E}[\widetilde{w}^{f}\mid \mathcal{G}]=\sum_{\ell \in
J_{+}}(q_{\ell }^{\top }\widetilde{w}^{f})q_{\ell },
\end{equation*}%
with $\operatorname{Var}(U^{\ast })=\sum_{\ell \in J_{+}}q_{\ell }q_{\ell }^{\top
}=V^{\ast }$. Thus the tie-broken optimum is attained by revealing exactly
the positive-eigenvalue coordinates: 
\begin{equation*}
r^{p}=\left( q_{\ell }^{\top }\widetilde{w}^{f}(X)\right) _{\ell \in J_{+}}.
\end{equation*}%
Equivalently, as a rating rule based on output, 
\begin{equation*}
r^{p}(Y)\equiv Q_{+}^{\top }\Sigma _{w^{f}}^{-1/2}w^{f}(Y-a^{p}).
\end{equation*}

Finally, for part (iii), since $r^{p}=Q_{+}^{\top }\widetilde{w}^{f}$ and $%
Q_{+}^{\top }Q_{+}=I_{|J_{+}|}$, 
\begin{equation*}
\mathbb{E}[\widetilde{w}^{f}\mid r^{p}]=Q_{+}r^{p},
\end{equation*}%
and therefore 
\begin{equation*}
w^{p}=\mathbb{E}[w^{f}\mid r^{p}] =\Sigma _{w^{f}}^{1/2}Q_{+}r^{p}.
\end{equation*}%
For uniqueness, let $w$ be any tie-broken optimal public posterior and set $%
U\equiv \Sigma _{w^{f}}^{-1/2}w$. Then $\operatorname{Var}(U)=V^{\ast }$ by part
(i). Since $\mathbb{E}[\widetilde{w}^{f}\mid U]=U$ by Lemma~\ref{lem:BP}(i), 
$\operatorname{Cov}(\widetilde{w}^{f},U)=\operatorname{Var}(U)=V^{\ast }$. Using $(V^{\ast
})^{2}=V^{\ast }$, 
\begin{equation*}
\operatorname{Var}(V^{\ast }\widetilde{w}^{f}-U)=0.
\end{equation*}%
Thus $U=V^{\ast }\widetilde{w}^{f}=Q_{+}r^{p}$ almost surely, and $w=\Sigma
_{w^{f}}^{1/2}Q_{+}r^{p}=w^{p}$.
\end{proof}

\textbf{\hypertarget{rem:A2}{Remark A2.}} When $M$ is nonsingular, $w^{p}$
is also a regular Lagrangian solution in Proposition~\ref{prop:relaxedpublic}%
. Write $w^{p}=Aw^{f}$ with $A\equiv \Sigma _{w^{f}}^{1/2}Q_{+}Q_{+}^{\top
}\Sigma _{w^{f}}^{-1/2}$. The first-order condition in the proof of
Proposition~\ref{prop:relaxedpublic} requires a symmetric $\Omega =\Psi
+\Psi ^{\top }$ with $\Omega A+A^{\top }\Omega -\Omega =2B$. Whitening by $%
\Sigma _{w^{f}}^{1/2}$ and passing to the eigenbasis of $M$ gives 
\begin{equation*}
\widehat{\Omega }_{k\ell }(d_{k}+d_{\ell }-1)=2\lambda _{k}1_{\{k=\ell
\}},\qquad d_{k}\equiv 1_{\{\lambda _{k}>0\}}.
\end{equation*}%
If $M$ is nonsingular, $\widehat{\Omega }=2|\Lambda |\succ 0$ solves this
equation, so 
\begin{equation*}
\Omega =2\Sigma _{w^{f}}^{-1/2}Q|\Lambda |Q^{\top }\Sigma
_{w^{f}}^{-1/2}\succ 0
\end{equation*}%
and $\Psi ^{\top }=\Omega A-\Gamma \Sigma _{\eta }^{-1}$ constitute a
regular Lagrangian solution, which also satisfies (\ref{Proof_Prop_6_Riccati}%
).

\begin{proof}[Proof of Corollary~\protect\ref{cor:fullpublic}]
Since $\Sigma _{w^{f}}^{1/2}$ is symmetric and nonsingular, $M=\Sigma
_{w^{f}}^{1/2}B\Sigma _{w^{f}}^{1/2}$ is congruent to $B$. Thus, by
Sylvester's law of inertia, $M$ and $B$ have the same eigenvalue signs.
Then, it suffices to ensure $B\succ 0$ for full disclosure.

Decompose $\Gamma $ into its mean and dispersion, 
\begin{equation*}
\Gamma =\overline{\gamma }I+D\text{, \ \ }D\equiv \operatorname{diag}(\gamma _{1}-%
\overline{\gamma },\ldots ,\gamma _{n}-\overline{\gamma })\text{,}
\end{equation*}%
so that $B=\frac{1}{2}(\Gamma \Sigma _{\eta }^{-1}+\Sigma _{\eta
}^{-1}\Gamma )=\overline{\gamma }\Sigma _{\eta }^{-1}+\frac{1}{2}(D\Sigma
_{\eta }^{-1}+\Sigma _{\eta }^{-1}D)$.

For any unit vector $x$, symmetry of $\Sigma _{\eta }^{-1}$ and $D$ yields $%
x^{\top }Bx=\overline{\gamma }x^{\top }\Sigma _{\eta }^{-1}x+x^{\top
}D\Sigma _{\eta }^{-1}x$, and the Rayleigh quotient together with $|x^{\top
}D\Sigma _{\eta }^{-1}x|\leq \Vert D\Vert _{2}\Vert \Sigma _{\eta
}^{-1}\Vert _{2}$ imply 
\begin{equation*}
x^{\top }Bx\geq \overline{\gamma }\lambda _{\min }(\Sigma _{\eta
}^{-1})-\left( \max_{i}|\gamma _{i}-\overline{\gamma }|\right) \lambda
_{\max }(\Sigma _{\eta }^{-1})\text{.}
\end{equation*}%
The right side is strictly positive iff $\max_{i}|\gamma _{i}-\overline{%
\gamma }|/\overline{\gamma }<\lambda _{\min }(\Sigma _{\eta })/\lambda
_{\max }(\Sigma _{\eta })$, the hypothesis of the corollary. Hence $B\succ 0$%
, and full disclosure is strictly optimal by Proposition~\ref{prop:gauss}.
\end{proof}

\begin{proof}[Proof of Corollary~\protect\ref{cor:partial}]
Given $\Sigma _{\eta }=(1-\rho )I_{n}+\rho \mathbf{11}^{\top }$ and $0<\rho
<1$, the Sherman--Morrison formula gives 
\begin{equation*}
\Sigma _{\eta }^{-1}=aI_{n}+b\mathbf{11}^{\top }\text{, \ \ \ }a=\frac{1}{%
1-\rho }\text{, \ \ }b=-\frac{\rho }{(1-\rho )(1+(n-1)\rho )}\text{.}
\end{equation*}%
Thus, 
\begin{equation*}
B\equiv \frac{\Gamma \Sigma _{\eta }^{-1}+\Sigma _{\eta }^{-1}\Gamma }{2}%
=a\Gamma +\frac{b}{2}(\Gamma \mathbf{11}^{\top }+\mathbf{11}^{\top }\Gamma ).
\end{equation*}%
Now define 
\begin{equation*}
C\equiv \Gamma ^{-1/2}B\Gamma ^{-1/2}=aI_{n}+\frac{b}{2}\left( \iota
_{+}\iota _{-}^{\top }+\iota _{-}\iota _{+}^{\top }\right) \text{, \ \ }%
\iota _{+}\equiv \Gamma ^{1/2}\mathbf{1}\text{, \ \ }\iota _{-}\equiv \Gamma
^{-1/2}\mathbf{1}\text{.}
\end{equation*}%
Because $\Gamma ^{-1/2}$ is symmetric and nonsingular, $C$ is congruent to $%
B $ and hence shares its eigenvalue signs.

Next, because $\iota _{+}\iota _{-}^{\top }+\iota _{-}\iota _{+}^{\top }$
has rank at most two, $C$ has $n-2$ eigenvalues equal to $a>0$. The two
remaining eigenvectors lie in $\operatorname{span}\{\iota _{+},\iota _{-}\}$.
Furthermore, 
\begin{equation*}
\iota _{+}^{\top }\iota _{+}=\sum_{i=1}^{n}\gamma _{i},\qquad \iota
_{-}^{\top }\iota _{-}=\sum_{i=1}^{n}\gamma _{i}^{-1},\qquad \iota
_{+}^{\top }\iota _{-}=n\text{,}
\end{equation*}%
so the two eigenvalues of $\frac{1}{2}(\iota _{+}\iota _{-}^{\top }+\iota
_{-}\iota _{+}^{\top })$ are $\lambda _{\pm }=(n\pm H)/2$, writing $H$ for $%
H(\Gamma )$. Those of $C$ are 
\begin{equation*}
\mu _{\pm }=a+b\lambda _{\pm }=a+b\frac{n\pm H}{2}\text{.}
\end{equation*}%
Since $a>0$, $b<0$, and $H\geq n$ by Cauchy--Schwarz, $\mu _{-}>0$. Then, $C$%
, and hence $B$, has exactly one nonpositive eigenvalue if and only if $\mu
_{+}\leq 0$; that is, 
\begin{equation*}
\frac{1}{1-\rho }-\frac{\rho }{(1-\rho )(1+(n-1)\rho )}\frac{n+H}{2}\leq 0.
\end{equation*}%
Rearranging terms reveals 
\begin{equation*}
\rho \geq \frac{2}{H-n+2}\equiv \rho ^{\ast }.
\end{equation*}%
Since $M$ and $B$ have the same eigenvalue signs, the corollary follows from
Proposition~\ref{prop:gauss}.
\end{proof}

\bigskip

In the following lemma, part (i) is the multivariate Gaussian version of
Tweedie's formula; see Robbins (\hyperlink{cite.Robbins1956}{%
\textcolor{linkblue}{1956}}) and Efron (\hyperlink{cite.Efron2011}{%
\textcolor{linkblue}{2011}}) for the scalar formulation. Part (ii) records a
sharpness observation: applied to Proposition~\ref{prop:noise} with $\xi
=\eta $ and $\theta =\varepsilon $, it says that Gaussian talent cannot be
dispensed with if the identity is to hold for every noise
distribution.\medskip

\noindent \textbf{\hypertarget{lem:A2}{Lemma A2.}} \textit{Let }$\theta \in 
\mathbb{R}^{n}$\textit{\ satisfy }$\mathbb{E}\Vert \theta \Vert <\infty $%
\textit{, and let }$\xi $\textit{\ be independent of }$\theta $\textit{,
with }$\mathbb{E}[\xi ]=0$\textit{, }$\operatorname{Var}(\xi )=\Sigma \succ 0$%
\textit{, and strictly positive }$C^{1}$\textit{\ density }$g$\textit{. Let }%
$Z=\theta +\xi $\textit{\ have marginal density }$p$\textit{. Then:}

\textit{(i) If }$\xi \sim N(0,\Sigma )$\textit{, then }$\mathbb{E}[\theta
\mid Z]=Z+\Sigma \nabla \ln p(Z)$\textit{\ almost surely.}

\textit{(ii) Conversely, if the identity in (i) holds for every distribution
of }$\theta $\textit{\ with }$\mathbb{E}\Vert \theta \Vert <\infty $\textit{%
, then }$\xi \sim N(0,\Sigma )$\textit{.\medskip }

\begin{proof}
For part (i), let $g$ be the $N(0,\Sigma )$ density, so that $\nabla
g(x)=-\Sigma ^{-1}x\,g(x)$. The marginal density of $Z$ is $p(z)=\int
g(z-\theta )dF(\theta )$, and differentiating under the integral sign, which 
$\mathbb{E}\Vert \theta \Vert <\infty $ and the Gaussian tails permit, 
\begin{equation*}
\nabla p(z)=\int \nabla g(z-\theta )dF(\theta )=-\Sigma ^{-1}\int (z-\theta
)g(z-\theta )dF(\theta )\text{.}
\end{equation*}%
Dividing by $p(z)$ and using $\nabla \ln p=\nabla p/p$, 
\begin{equation*}
\nabla \ln p(z)=-\Sigma ^{-1}\mathbb{E}[Z-\theta \mid Z=z]=-\Sigma
^{-1}\left( z-\mathbb{E}[\theta \mid Z=z]\right) \text{,}
\end{equation*}%
and premultiplying by $\Sigma $ proves part (i).

For part (ii), take $\theta $ degenerate at an arbitrary $\theta _{0}\in 
\mathbb{R}^{n}$, which is admissible since $\mathbb{E}\Vert \theta \Vert
<\infty $. Then $p(z)=g(z-\theta _{0})$ and $\mathbb{E}[\theta \mid
Z]=\theta _{0}$, so the identity reads $\theta _{0}=z+\Sigma \nabla \ln
g(z-\theta _{0})$ for every $z$; writing $x\equiv z-\theta _{0}$, this is $%
\nabla \ln g(x)=-\Sigma ^{-1}x$ for every $x$. Since $g$ is strictly
positive and $C^{1}$, integrating gives $\ln g(x)=-\frac{1}{2}x^{\top
}\Sigma ^{-1}x+c$, so $g$ is the $N(0,\Sigma )$ density.
\end{proof}

\begin{proof}[Proof of Proposition~\protect\ref{prop:noise}]
Applying Lemma~\hyperlink{lem:A2}{A2} with $\xi =\eta \sim N(0,\Sigma _{\eta
})$, $\theta =\varepsilon $, and $Z=X$ yields $\mathbb{E}[\varepsilon \mid
X]=X+\Sigma _{\eta }\nabla \ln f_{X}(X)$. Since $w^{f}(X)=X-\mathbb{E}%
[\varepsilon \mid X]$ and $s(X)=-\nabla \ln f_{X}(X)$, 
\begin{equation*}
s(X)=\Sigma _{\eta }^{-1}w^{f}(X)\text{,}
\end{equation*}%
whatever the distribution of $\varepsilon $, so long as $\mathbb{E}\Vert
\varepsilon \Vert <\infty $. Hence (\ref{Public_Effort}) applies to the
posterior $w$ induced by any public rating.

Let $\zeta \equiv w^{f}-w$. Bayes plausibility, $\mathbb{E}[w^{f}\mid w]=w$,
gives $\mathbb{E}[\zeta ]=0$ and $\operatorname{Cov}(w,\zeta )=0$, so $\operatorname{Var}%
(w^{f})=\operatorname{Var}(w)+\operatorname{Var}(\zeta )$ and 
\begin{equation*}
\mathcal{E}=\mathcal{E}^{f}-\operatorname{tr}\left( B\operatorname{Var}(\zeta )\right) \text{%
.}
\end{equation*}

If $B\succ 0$ and $\zeta \neq 0$ with positive probability, then $\operatorname{Var}%
(\zeta )$ is nonzero and positive semidefinite, so $\operatorname{tr}(B\operatorname{Var}%
(\zeta ))\geq \lambda _{\min }(B)\operatorname{tr}(\operatorname{Var}(\zeta ))>0$ and $%
\mathcal{E}<\mathcal{E}^{f}$. Full disclosure, which implements $\zeta =0$,
is thus strictly optimal.

Conversely, suppose $B\nsucc 0$, and let $d$ be a unit eigenvector of $B$
with eigenvalue $\lambda \leq 0$. Disclose $r\equiv (I_{n}-dd^{\top })w^{f}$%
, so that $w=\mathbb{E}[w^{f}\mid r]$ and $(I_{n}-dd^{\top })\zeta =0$. Then 
$\operatorname{Var}(\zeta )=c\,dd^{\top }$ with $c=\mathbb{E}[\operatorname{Var}(d^{\top
}w^{f}\mid r)]>0$, since absolute continuity of $w^{f}$ makes $d^{\top
}w^{f} $ non-$r$-measurable. Hence $\mathcal{E}=\mathcal{E}^{f}-c\lambda
\geq \mathcal{E}^{f}$, and full disclosure is not strictly optimal.
\end{proof}

\begin{proof}[Proof of Corollary~\protect\ref{cor:spillconf}]
Replacing $s_{i}$ by $\widetilde{s}_{i}$ in Proposition~\ref{prop:relaxed}, $%
w_{i}^{o}=\frac{1}{2}[w_{i}^{f}+\kappa _{i}(\beta )\widetilde{s}_{i}]$ with $%
\kappa _{i}(\beta )=[\operatorname{Var}(w_{i}^{f})/\operatorname{Var}(\widetilde{s}%
_{i})]^{1/2}$. Since $w_{i}^{o}$ is linear in $X$, summing its peer weights
and using (\ref{deriv_cov}) yields, with $S_{-i}\equiv \sum_{j\neq i}s_{j}$, 
\begin{equation*}
T_{i}(\beta )=\frac{1}{2}\left[ \operatorname{Cov}(w_{i}^{f},S_{-i})+\operatorname{Var}%
(w_{i}^{f})^{1/2}\frac{u(\beta )}{v(\beta )^{1/2}}\right] \text{, \ \ }%
u\equiv \operatorname{Cov}(\widetilde{s}_{i},S_{-i})\text{, \ }v\equiv \operatorname{Var}(%
\widetilde{s}_{i})\text{.}
\end{equation*}%
Only $u/\sqrt{v}$ varies with $\beta $. By (\ref{score_spillover}), $%
\widetilde{s}_{i}=s_{i}+\delta S_{-i}$ with $\delta \equiv \beta /(n-1)$, so 
$du/d\delta =\operatorname{Var}(S_{-i})$ and $dv/d\delta =2u$, giving $\frac{d}{%
d\delta }(uv^{-1/2})=\left[ \operatorname{Var}(S_{-i})v-u^{2}\right] v^{-3/2}$.
Substituting, the terms in $\delta $ cancel and 
\begin{equation*}
\operatorname{Var}(S_{-i})v-u^{2}=\operatorname{Var}(s_{i})\operatorname{Var}(S_{-i})-\operatorname{Cov}%
(s_{i},S_{-i})^{2}>0
\end{equation*}%
by Cauchy--Schwarz, strictly because $\Sigma _{X}\succ 0$ keeps $s_{i}$ and $%
S_{-i}$ from being proportional. Since $d\delta /d\beta =1/(n-1)>0$, $%
T_{i}^{\prime }(\beta )>0$.
\end{proof}

\begin{proof}[Proof of Proposition~\protect\ref{prop:spill}]
The common factor $1+\beta >0$ scales total output and can be omitted in the
objective, as can $\gamma $. Let $\delta \equiv \beta /(n-1)$. By (\ref%
{score_spillover}), agent $i$'s score is the $i$th component of $G_{\beta
}s(X)$, where $G_{\beta }\equiv (1-\delta )I_{n}+\delta \mathbf{1}\mathbf{1}%
^{\top }$, so the objective in ($\mathcal{P}2$) becomes $\mathbb{E}[w^{\top
}G_{\beta }s(X)]$. As in the no-spillover case, with $s=\Sigma _{\eta
}^{-1}w^{f}$ under normality, the objective becomes the quadratic form: 
\begin{equation*}
\mathbb{E}[w^{\top }B_{\beta }w]\text{, \ \ \ }B_{\beta }\equiv \tfrac{1}{2}%
\left( G_{\beta }\Sigma _{\eta }^{-1}+\Sigma _{\eta }^{-1}G_{\beta }\right) 
\text{.}
\end{equation*}%
The remainder parallels the proof of Corollary~\ref{cor:partial}. Congruence
by $\Sigma _{\eta }^{1/2}$ produces 
\begin{equation*}
C_{\beta }\equiv \Sigma _{\eta }^{1/2}B_{\beta }\Sigma _{\eta
}^{1/2}=(1-\delta )I_{n}+\tfrac{\delta }{2}(\iota _{+}\iota _{-}^{\top
}+\iota _{-}\iota _{+}^{\top })\text{,\quad }\iota _{+}\equiv \Sigma _{\eta
}^{1/2}\mathbf{1}\text{,\quad }\iota _{-}\equiv \Sigma _{\eta }^{-1/2}%
\mathbf{1}\text{,}
\end{equation*}%
which has the form of $C$ there with $(a,b)=(1-\delta ,\delta )$ and $\iota
_{+}^{\top }\iota _{+}=\mathbf{1}^{\top }\Sigma _{\eta }\mathbf{1}$, $\iota
_{-}^{\top }\iota _{-}=\mathbf{1}^{\top }\Sigma _{\eta }^{-1}\mathbf{1}$, $%
\iota _{+}^{\top }\iota _{-}=n$. The same argument shows that $C_{\beta }$
has $n-2$ eigenvalues equal to $1-\delta $, and the remaining two are 
\begin{equation*}
\mu _{\pm }=1-\delta +\delta \frac{n\pm H(\Sigma _{\eta })}{2}\text{.}
\end{equation*}%
Cauchy--Schwarz applied to $\iota _{+}$ and $\iota _{-}$, together with $%
\iota _{+}^{\top }\iota _{-}=n$, yields $H(\Sigma _{\eta })\geq n$. Since $%
|\beta |<1$, $1-\delta >0$. Moreover, 
\begin{equation*}
\mu _{+}=1+\frac{\delta }{2}\left( H(\Sigma _{\eta })+n-2\right) \text{,
\qquad }\mu _{-}=1-\frac{\delta }{2}\left( H(\Sigma _{\eta })-n+2\right) 
\text{.}
\end{equation*}%
If $\beta \geq 0$, then $\mu _{+}\geq 1+\beta >0$, while $\mu _{-}>0$ if and
only if $\beta <\beta ^{\ast }=\frac{2(n-1)}{H(\Sigma _{\eta })-n+2}$. If $%
\beta <0$, then $\mu _{-}>1$, while $\mu _{+}>0$ if and only if $\beta
>\beta _{\ast }=-\frac{2(n-1)}{H(\Sigma _{\eta })+n-2}$. That $H(\Sigma
_{\eta })\geq n$ also implies $\beta ^{\ast }\leq n-1$ and $\beta _{\ast
}\geq -1$, with equality in each if and only if $H(\Sigma _{\eta })=n$.

Finally, $M\equiv \Sigma _{w^{f}}^{1/2}B_{\beta }\Sigma _{w^{f}}^{1/2}$ is
congruent to $B_{\beta }$, hence to $C_{\beta }$, so $|J_{+}|$ counts the
positive eigenvalues of $C_{\beta }$. For $\beta _{\ast }<\beta <\beta
^{\ast }$, all $n$ are positive. For $\beta \leq \beta _{\ast }$, only $\mu
_{+}$ is nonpositive, and for $\beta ^{\ast }\leq \beta <1$, only $\mu _{-}$
is, leaving $n-1$ in both cases. Proposition~\ref{prop:gauss}(ii) then
delivers the dimension, and full disclosure is strictly optimal exactly when 
$\beta _{\ast }<\beta <\beta ^{\ast }$.
\end{proof}


\newpage

\begin{thebibliography}{99}
\bibitem{AoyagiMenuet2026}\hypertarget{cite.AoyagiMenuet2026}{} \noindent Aoyagi, Masaki, and Maxime Menuet.
2026. `` Incentive Compatible Information
Disclosure.''\ Working paper.

\bibitem{ArieliBabichenko2019}\hypertarget{cite.ArieliBabichenko2019}{} \noindent Arieli, Itai, and Yakov Babichenko.
2019. `` Private Bayesian Persuasion.''\ 
\textit{Journal of Economic Theory} 182: 185--217.

\bibitem{AryaMittendorf2011}\hypertarget{cite.AryaMittendorf2011}{} \noindent Arya, Anil, and Brian Mittendorf.
2011. `` The Benefits of Aggregate Performance Metrics in the
Presence of Career Concerns.''\ \textit{Management Science}
57 (8): 1424--1437.

\bibitem{AuriolEtAl2002}\hypertarget{cite.AuriolEtAl2002}{} \noindent Auriol, Emmanuelle, Guido Friebel, and
Lambros Pechlivanos. 2002. `` Career Concerns in
Teams.''\ \textit{Journal of Labor Economics} 20 (2):
289--307.

\bibitem{BardhiGuo2018}\hypertarget{cite.BardhiGuo2018}{} \noindent Bardhi, Arjada, and Yingni Guo. 2018.
`` Modes of Persuasion toward Unanimous
Consent.''\ \textit{Theoretical Economics} 13 (3):
1111--1149.

\bibitem{BarIsaac2007}\hypertarget{cite.BarIsaac2007}{} \noindent Bar-Isaac, Heski. 2007. ``
Something to Prove: Reputation in Teams.''\ \textit{RAND
Journal of Economics} 38 (2): 495--511.

\bibitem{BarIsaacDeb2021}\hypertarget{cite.BarIsaacDeb2021}{} \noindent Bar-Isaac, Heski, and Joyee Deb. 2021.
`` Reputation with Opportunities for
Coasting.''\ \textit{Journal of the European Economic
Association} 19 (1): 200--236.

\bibitem{Bishop2013}\hypertarget{cite.Bishop2013}{} \noindent Bishop, Todd. 2013. ``
Microsoft Axes `Stack Ranking' in Massive Overhaul of Employee Review
Process.''\ \textit{GeekWire}, November 12.

\bibitem{BuckinghamGoodall2015}\hypertarget{cite.BuckinghamGoodall2015}{} \noindent Buckingham, Marcus, and Ashley
Goodall. 2015. `` Reinventing Performance
Management.''\ \textit{Harvard Business Review} 93 (4):
40--50.

\bibitem{BuffaEtAl2025}\hypertarget{cite.BuffaEtAl2025}{} \noindent Buffa, Andrea M., Qing Liu, and Lucy
White. 2025. `` Privacy and Team
Incentives.''\ \textit{Journal of Finance} 80 (6):
3443--3498.

\bibitem{Chalioti2016}\hypertarget{cite.Chalioti2016}{} \noindent Chalioti, Evangelia. 2016.
`` Team Production, Endogenous Learning about Abilities, and
Career Concerns.''\ \textit{European Economic Review} 85:
229--244.

\bibitem{ChanEtAl2019}\hypertarget{cite.ChanEtAl2019}{} \noindent Chan, Jimmy, Seher Gupta, Fei Li, and Yun
Wang. 2019. `` Pivotal Persuasion.''\ \textit{%
Journal of Economic Theory} 180: 178--202.

\bibitem{Cohan2013}\hypertarget{cite.Cohan2013}{} \noindent Cohan, Peter. 2013. `` Adobe's
Stock Up $68\%$ Since It Dumped Stack Ranking, Will Microsoft's
Follow?''\ \textit{Forbes}, November 29.

\bibitem{CullenPakzadHurson2023}\hypertarget{cite.CullenPakzadHurson2023}{} \noindent Cullen, Zo\"{e} B., and Bobak
Pakzad-Hurson. 2023. `` Equilibrium Effects of Pay
Transparency.''\ \textit{Econometrica} 91 (3): 765--802.

\bibitem{DewatripontEtAl1999a}\hypertarget{cite.DewatripontEtAl1999a}{} \noindent Dewatripont, Mathias, Ian Jewitt,
and Jean Tirole. 1999a. `` The Economics of Career Concerns,
Part I: Comparing Information Structures.''\ \textit{Review
of Economic Studies} 66 (1): 183--198.

\bibitem{DewatripontEtAl1999b}\hypertarget{cite.DewatripontEtAl1999b}{} \noindent Dewatripont, Mathias, Ian Jewitt,
and Jean Tirole. 1999b. `` The Economics of Career Concerns,
Part II: Application to Missions and Accountability of Government
Agencies.''\ \textit{Review of Economic Studies} 66 (1):
199--217.

\bibitem{DworczakKolotilin2024}\hypertarget{cite.DworczakKolotilin2024}{} \noindent Dworczak, Piotr, and Anton
Kolotilin. 2024. `` The Persuasion
Duality.''\ \textit{Theoretical Economics} 19 (4):
1701--1755.

\bibitem{Efron2011}\hypertarget{cite.Efron2011}{} \noindent Efron, Bradley. 2011. ``
Tweedie's Formula and Selection Bias.''\ \textit{Journal of
the American Statistical Association} 106 (496): 1602--1614.

\bibitem{Eichenwald2012}\hypertarget{cite.Eichenwald2012}{} \noindent Eichenwald, Kurt. 2012. ``
Microsoft's Lost Decade.''\ \textit{Vanity Fair} 54 (8).

\bibitem{Ekmekci2011}\hypertarget{cite.Ekmekci2011}{} \noindent Ekmekci, Mehmet. 2011. ``
Sustainable Reputations with Rating Systems.''\ \textit{%
Journal of Economic Theory} 146 (2): 479--503.

\bibitem{Elias2022}\hypertarget{cite.Elias2022}{} \noindent Elias, Jennifer. 2022. ``
Google Tells Employees More of Them Will Be at Risk for Low Performance
Ratings Next Year.''\ \textit{CNBC}, December 22.

\bibitem{Fedweek2019}\hypertarget{cite.Fedweek2019}{} \noindent Fedweek. 2019. `` OPM Sets
New Standards on Performance Ratings.''\ July 17.

\bibitem{FleckingerEtAl2024}\hypertarget{cite.FleckingerEtAl2024}{} \noindent Fleckinger, Pierre, David Martimort,
and Nicolas Roux. 2024. `` Should They Compete or Should They
Cooperate? The View of Agency Theory.''\ \textit{Journal of
Economic Literature} 62 (4): 1589--1646.

\bibitem{GanuzaPenalva2010}\hypertarget{cite.GanuzaPenalva2010}{} \noindent Ganuza, Juan-Jos\'{e}, and Jos\'{e} S.
Penalva. 2010. `` Signal Orderings Based on Dispersion and
the Supply of Private Information in Auctions.''\ \textit{%
Econometrica} 78 (3): 1007--1030.

\bibitem{GoldmanisRay2021}\hypertarget{cite.GoldmanisRay2021}{} \noindent Goldmanis, Maris, and Korok Ray. 2021.
`` Team Incentives under Private
Contracting.''\ \textit{RAND Journal of Economics} 52 (2):
334--358.

\bibitem{GreenStokey1983}\hypertarget{cite.GreenStokey1983}{} \noindent Green, Jerry R., and Nancy L. Stokey.
1983. `` A Comparison of Tournaments and
Contracts.''\ \textit{Journal of Political Economy} 91 (3):
349--364.

\bibitem{HalacEtAl2021}\hypertarget{cite.HalacEtAl2021}{} \noindent Halac, Marina, Elliot Lipnowski, and
Daniel Rappoport. 2021. `` Rank Uncertainty in
Organizations.''\ \textit{American Economic Review} 111 (3):
757--786.

\bibitem{Hansen2013}\hypertarget{cite.Hansen2013}{} \noindent Hansen, Stephen E. 2013. ``
Performance Feedback with Career Concerns.''\ \textit{%
Journal of Law, Economics, and Organization} 29 (6): 1279--1316.

\bibitem{Holmstrom1982}\hypertarget{cite.Holmstrom1982}{} \noindent Holmstr\"{o}m, Bengt. 1982.
`` Moral Hazard in Teams.''\ \textit{Bell
Journal of Economics} 13 (2): 324--340.

\bibitem{Holmstrom1999}\hypertarget{cite.Holmstrom1999}{} \noindent Holmstr\"{o}m, Bengt. 1999.
`` Managerial Incentive Problems: A Dynamic
Perspective.''\ \textit{Review of Economic Studies} 66 (1):
169--182.

\bibitem{HornerLambert2021}\hypertarget{cite.HornerLambert2021}{} \noindent H\"{o}rner, Johannes, and Nicolas S.
Lambert. 2021. `` Motivational Ratings.''\ 
\textit{Review of Economic Studies} 88 (4): 1892--1935.

\bibitem{Kamenica2019}\hypertarget{cite.Kamenica2019}{} \noindent Kamenica, Emir. 2019. ``
Bayesian Persuasion and Information Design.''\ \textit{%
Annual Review of Economics} 11: 249--272.

\bibitem{KamenicaGentzkow2011}\hypertarget{cite.KamenicaGentzkow2011}{} \noindent Kamenica, Emir, and Matthew
Gentzkow. 2011. `` Bayesian Persuasion.''\ 
\textit{American Economic Review} 101 (6): 2590--2615.

\bibitem{KamienEtAl1992}\hypertarget{cite.KamienEtAl1992}{} \noindent Kamien, Morton I., Eitan Muller, and
Israel Zang. 1992. `` Research Joint Ventures and R\&D
Cartels.''\ \textit{American Economic Review} 82 (5):
1293--1306.

\bibitem{KarlinRinott1980}\hypertarget{cite.KarlinRinott1980}{} \noindent Karlin, Samuel, and Yosef Rinott. 1980.
`` Classes of Orderings of Measures and Related Correlation
Inequalities. I. Multivariate Totally Positive
Distributions.''\ \textit{Journal of Multivariate Analysis}
10 (4): 467--498.

\bibitem{Kim2024}\hypertarget{cite.Kim2024}{} \noindent Kim, Eugene. 2024. `` Official
Amazon HR Document Reveals a Spike in Employees Put on PIPs During a Time of
Record Layoffs.''\ \textit{Business Insider}, March 20.

\bibitem{LazearRosen1981}\hypertarget{cite.LazearRosen1981}{} \noindent Lazear, Edward P., and Sherwin Rosen.
1981. `` Rank-Order Tournaments as Optimum Labor
Contracts.''\ \textit{Journal of Political Economy} 89 (5):
841--864.

\bibitem{LinEtAl2024}\hypertarget{cite.LinEtAl2024}{} \noindent Lin, Luona, Juliana Menasce Horowitz, and
Richard Fry. 2024. `` Most Americans Feel Good About Their
Job Security but Not Their Pay.''\ \textit{Pew Research
Center}, December 10.

\bibitem{MacLeodEtAl2017}\hypertarget{cite.MacLeodEtAl2017}{} \noindent MacLeod, W. Bentley, Evan Riehl, Juan E.
Saavedra, and Miguel Urquiola. 2017. `` The Big Sort: College
Reputation and Labor Market Outcomes.''\ \textit{American
Economic Journal: Applied Economics} 9 (3): 223--261.

\bibitem{MeagherPrasad2016}\hypertarget{cite.MeagherPrasad2016}{} \noindent Meagher, Kieron, and Suraj Prasad.
2016. `` Career Concerns and Team Talent.''\ 
\textit{Journal of Economic Behavior and Organization} 129: 1--17.

\bibitem{Meyer1994}\hypertarget{cite.Meyer1994}{} \noindent Meyer, Margaret A. 1994. `` The
Dynamics of Learning with Team Production: Implications for Task
Assignment.''\ \textit{Quarterly Journal of Economics} 109
(4): 1157--1184.

\bibitem{MeyerVickers1997}\hypertarget{cite.MeyerVickers1997}{} \noindent Meyer, Margaret A., and John Vickers.
1997. `` Performance Comparisons and Dynamic
Incentives.''\ \textit{Journal of Political Economy} 105
(3): 547--581.

\bibitem{NalebuffStiglitz1983}\hypertarget{cite.NalebuffStiglitz1983}{} \noindent Nalebuff, Barry J., and Joseph E.
Stiglitz. 1983. `` Prizes and Incentives: Towards a General
Theory of Compensation and Competition.''\ \textit{Bell
Journal of Economics} 14 (1): 21--43.

\bibitem{NeemanEtAl2019}\hypertarget{cite.NeemanEtAl2019}{} \noindent Neeman, Zvika, Anik\'{o} \"{O}ry, and
Jungju Yu. 2019. `` The Benefit of Collective
Reputation.''\ \textit{RAND Journal of Economics} 50 (4):
787--821.

\bibitem{OPM2025}\hypertarget{cite.OPM2025}{} \noindent Office of Personnel Management. 2025.
`` Assuring Responsive and Accountable Federal Executive
Management.''\ Final rule, 90 Fed.\ Reg.\ 44299, September
15.

\bibitem{OPM2026}\hypertarget{cite.OPM2026}{} \noindent Office of Personnel Management. 2026.
`` Performance Appraisal for General Schedule, Prevailing
Rate, and Certain Other Employees.''\ Final rule, 91 Fed.\
Reg.\ 41521, July 7.

\bibitem{Ortega2003}\hypertarget{cite.Ortega2003}{} \noindent Ortega, Jaime. 2003. `` Power
in the Firm and Managerial Career Concerns.''\ \textit{%
Journal of Economics \& Management Strategy} 12 (1): 1--29.

\bibitem{Pei2016}\hypertarget{cite.Pei2016}{} \noindent Pei, Harry Di. 2016. ``
Reputation with Strategic Information Disclosure.''\ Working
paper, Massachusetts Institute of Technology.

\bibitem{RayoSegal2010}\hypertarget{cite.RayoSegal2010}{} \noindent Rayo, Luis, and Ilya Segal. 2010.
`` Optimal Information Disclosure.''\ \textit{%
Journal of Political Economy} 118 (5): 949--987.

\bibitem{Robbins1956}\hypertarget{cite.Robbins1956}{} \noindent Robbins, Herbert. 1956. `` An
Empirical Bayes Approach to Statistics.''\ In \textit{%
Proceedings of the Third Berkeley Symposium on Mathematical Statistics and
Probability}, vol. 1, 157--163. Berkeley: University of California Press.

\bibitem{Rodina2017}\hypertarget{cite.Rodina2017}{} \noindent Rodina, David. 2017. ``
Information Design and Career Concerns.''\ Working paper,
University of Bonn.

\bibitem{RodriguezHorwitz2023}\hypertarget{cite.RodriguezHorwitz2023}{} \noindent Rodriguez, Salvador, and Jeff
Horwitz. 2023. `` Facebook Parent Meta Gives Thousands of
Workers Subpar Reviews.''\ \textit{Wall Street Journal},
February 17.

\bibitem{Rosenfeld2017}\hypertarget{cite.Rosenfeld2017}{} \noindent Rosenfeld, Jake. 2017. ``
Don't Ask or Tell: Pay Secrecy Policies in U.S.
Workplaces.''\ \textit{Social Science Research} 65: 1--16.

\bibitem{Smolin2021}\hypertarget{cite.Smolin2021}{} \noindent Smolin, Alex. 2021. `` Dynamic
Evaluation Design.''\ \textit{American Economic Journal:
Microeconomics} 13 (4): 300--331.

\bibitem{WijayantiEtAl2025}\hypertarget{cite.WijayantiEtAl2025}{} \noindent Wijayanti, Aniek, Mahfud Sholihin,
Ertambang Nahartyo, and Supriyadi. 2025. `` What Do We Know
about the Forced Distribution System? A Systematic Literature Review and
Opportunities for Future Research.''\ \textit{Management
Review Quarterly} 75 (1): 747--788.

\bibitem{Yildirim2023}\hypertarget{cite.Yildirim2023}{} \noindent Yildirim, Huseyin. 2023. ``
Who Fares Better in Teamwork?''\ \textit{RAND Journal of
Economics} 54 (2): 299--324.

\bibitem{Yildirim2025}\hypertarget{cite.Yildirim2025}{} \noindent Yildirim, Huseyin. 2025. ``
The Economics of Career Concerns in Teamwork.''\ \textit{%
Journal of Economic Theory} 226: 106017.
\end{thebibliography}
\end{document}